\documentclass[journal]{IEEEtran}
\usepackage{amsmath,amsfonts}
\usepackage{algorithmic}
\usepackage{algorithm}
\usepackage{array}
\usepackage[caption=false,font=normalsize,labelfont=sf,textfont=sf]{subfig}
\usepackage{textcomp}
\usepackage{stfloats}
\usepackage{color,soul}
\usepackage[hyphens]{url}
\usepackage{pifont}
\usepackage{multirow}
\usepackage{verbatim}
\usepackage{graphicx}
\usepackage{cite}
\usepackage{setspace}
\usepackage[hidelinks]{hyperref}
\usepackage{scalerel}
\usepackage{tikz}
\usetikzlibrary{svg.path}

\definecolor{orcidlogocol}{HTML}{A6CE39}
\tikzset{
	orcidlogo/.pic={
		\fill[orcidlogocol] svg{M256,128c0,70.7-57.3,128-128,128C57.3,256,0,198.7,0,128C0,57.3,57.3,0,128,0C198.7,0,256,57.3,256,128z};
		\fill[white] svg{M86.3,186.2H70.9V79.1h15.4v48.4V186.2z}
		svg{M108.9,79.1h41.6c39.6,0,57,28.3,57,53.6c0,27.5-21.5,53.6-56.8,53.6h-41.8V79.1z M124.3,172.4h24.5c34.9,0,42.9-26.5,42.9-39.7c0-21.5-13.7-39.7-43.7-39.7h-23.7V172.4z}
		svg{M88.7,56.8c0,5.5-4.5,10.1-10.1,10.1c-5.6,0-10.1-4.6-10.1-10.1c0-5.6,4.5-10.1,10.1-10.1C84.2,46.7,88.7,51.3,88.7,56.8z};
	}
}
\newcommand\orcidicon[1]{$^{\href{https://orcid.org/#1}{\mbox{\scalerel*{
				\begin{tikzpicture}[yscale=-1,transform shape]
				\pic{orcidlogo};
				\end{tikzpicture}
			}{i}}}}$}
   
\definecolor{dgreen}{rgb}{0.00, 0.5, 0.00}
\def\BibTeX{{\rm B\kern-.05em{\sc i\kern-.025em b}\kern-.08em
    T\kern-.1667em\lower.7ex\hbox{E}\kern-.125emX}}

\newcommand{\cmmnt}[1]{\ignorespaces}
\newcommand{\mr}[2][2]{\multirow{#1}{*}{#2}}

\definecolor{lightyellow}{RGB}{252,244,133} \definecolor{darkyellow}{RGB}{255,209,0}
\definecolor{lightRed}{RGB}{242,220,218} \definecolor{darkRed}{RGB}{217,150,144}
\definecolor{lightPurple}{RGB}{221,215,230} \definecolor{darkPurple}{RGB}{205,194,217}
\definecolor{lightGreen}{RGB}{235,241,223} \definecolor{darkGreen}{RGB}{205,220,175}
\definecolor{lightBlue}{RGB}{219,238,244} \definecolor{darkBlue}{RGB}{147,205,221}

\DeclareRobustCommand{\ronenew}[1]{{#1}}

\DeclareRobustCommand{\rtwonew}[1]{{#1}}
\newcommand{\hll}[1]{{#1}}

\title{CiMBA: Accelerating Genome Sequencing through On-Device Basecalling via Compute-in-Memory}
\author{William Andrew Simon\orcidicon{0000-0001-7357-7204}, Irem~Boybat\orcidicon{0000-0002-4255-8622}, Riselda Kodra\orcidicon{0009-0001-6998-0698}, Elena Ferro\orcidicon{0000-0002-8618-8643},\\ Gagandeep Singh\orcidicon{0000-0002-3502-7401}, Mohammed Alser\orcidicon{0000-0002-6117-3701}, Shubham Jain\orcidicon{0000-0002-2291-7712}, Hsinyu Tsai\orcidicon{0000-0002-3971-097X
}~\IEEEmembership{Senior Member,~IEEE,}\\ Geoffrey W. Burr\orcidicon{0000-0001-5717-2549}~\IEEEmembership{Fellow,~IEEE,} Onur Mutlu\orcidicon{0000-0002-0075-2312}~\IEEEmembership{Fellow,~IEEE,} Abu Sebastian\orcidicon{0000-0001-5603-5243}~\IEEEmembership{Fellow,~IEEE}

\IEEEcompsocitemizethanks{ 
\IEEEcompsocthanksitem This work was supported by European Union’s Horizon Europe Research and Innovation Program (BioPIM, Grant 101047160), and Swiss State Secretariat for Education, Research and Innovation (SERI) (Grant 22.00076).
\IEEEcompsocthanksitem All authors except G. Singh are IEEE Members. R. Kodra and E. Ferro are Student Members.
\IEEEcompsocthanksitem WA. Simon, I. Boybat, E. Ferro, S. Jain, H. Tsai, G. Burr, and A. Sebastian are affiliated with International Business Machines (IBM).
\IEEEcompsocthanksitem R. Kodra was affiliated with IBM at time of writing and is currently affiliated with the Swiss Federal Institute of Technology, Lausanne.
\IEEEcompsocthanksitem G. Singh is associated Advanced Micro Devices (AMD).
\IEEEcompsocthanksitem M. Alser is associated with Georgia State University.
\IEEEcompsocthanksitem Onur Mutlu is affiliated with the ETH, Zurich. 
}
}

\begin{document}
\bstctlcite{IEEEexample:BSTcontrol}

\maketitle
\thispagestyle{plain}
\pagestyle{plain}

\begin{abstract}
As genome sequencing is finding utility in a wide variety of domains beyond the confines of traditional medical settings, its computational pipeline faces two significant challenges. First, the creation of up to 0.5 GB of data per minute imposes substantial communication and storage overheads. Second, the sequencing pipeline is bottlenecked at the basecalling step, consuming \textgreater40\% of genome analysis time. A range of proposals have attempted to address these challenges, with limited success. 

We propose to address these challenges with a Compute-in-Memory Basecalling Accelerator (CiMBA), the first embedded ($\sim25$mm$^2$)  accelerator capable of real-time, on-device basecalling, coupled with AnaLog (AL)-Dorado, a new family of analog focused basecalling DNNs. Our resulting hardware/software co-design greatly reduces data communication overhead, is capable of a throughput of 4.77 million bases per second, 24$\times$ that required for real-time operation, and achieves 17$\times$/27$\times$ power/area efficiency over the best prior basecalling embedded accelerator
while maintaining a high accuracy comparable to state-of-the-art software basecallers.
\end{abstract}


\section{Introduction}
\label{sec:intro}
\IEEEPARstart{A}{dvances} such as rapid genetic disease diagnosis~\cite{clark2019}, individually tailored precision therapies~\cite{ashley2016}, and preventive medicine~\cite{flores2013} \hll{have all been realized} in part due to plummeting costs as sequencing devices and applications mature~\cite{wetterstrand2021}. 
As cost barriers to entry have disappeared, personalized genomics has seen rapid uptake in both urban areas~\cite{sweeney2021} and rural communities~\cite{quick2016}, and across a wide range of life science applications such as forensics~\cite{borsting2015,maria2017}, and crop improvement~\cite{cruz2023}, with research occurring in increasingly remote areas, such as disused coalmines~\cite{SANG2018173} and outer space~\cite{mora2019}.

As the application range increases, genome sequencing's current limitations become more acute. In particular, Oxford Nanopore Technology's (ONT's) portable sequencing device, the MinION~\cite{wang2021}, is capable of producing up to 0.46 GB of sequencing data per minute~\cite{benton2022}. \hll{Transducing raw signal data into base nucleotide (e.g., A, C, G, or T) sequences involves a computationally expensive step called "basecalling".}

Modern basecalling algorithms incorporate Deep Neural Networks (DNNs), which improve accuracy yet can consume between 40\%~\cite{lou2020} (Tesla T4 GPU) and 86\%~\cite{bowden2019} (24 CPU threads) of the total execution time of genome analysis. 
The MinION, ONT's most portable device, lacks sufficient computational power to perform basecalling, necessitating constant connectivity to off-device storage and compute. The MinION Mk1C, with its integrated Jetson TX2 GPU, greatly increases sequencing portability. However, its compute power is barely sufficient for real-time basecalling~\cite{benton2022}, and thus risks being overwhelmed by expected improvements in flow cell technology~\cite{oxford2022}. 
In general, advances in sequencing capability have far outpaced available computational power~\cite{stephens2015}. Introducing more compute in the form of CPUs/GPUs can only partially solve the problem, given the time and energy needed to move such massive amounts of data over to this compute~\cite{oliveira2021,boroumand2021}. Cloud computing faces its own unique challenges in terms of privacy~\cite{alser2016,almadhoun2020_2} and security~\cite{almadhoun2020}. There is a need to process the raw data into nucleotide sequences in real-time, at the point of data generation, with high energy- and area efficiency.

Many proposed solutions attempt to address basecalling computational complexity~\cite{wu2019, wu2022, hammad2021, singh2024,lou2018, lou2020, weng2023, ferguson2022}, reduce its memory footprint~\cite{gamaarachchi2022, kovaka2023}, further accelerate it via GPU~\cite{wick2019, singh2024, goenka2022, xu2021,lv2020, zeng2020, yeh2022}, FPGA~\cite{wu2019, hammad2021}, TPU~\cite{perešíni2021}, or spatial architectures such as AMD-Xilinx's Versal AI Engine~\cite{singh2024} or in-memory computing~\cite{lou2018, lou2020, mao2022, shahroodi2023swordfish}. Other works have focused on utilizing ONT's "read-until" feature, which allows "unwanted" reads to be terminated mid-sequence~\cite{loose2016, kovaka2021, dunn2021,2021zhang, firtina2023, cavlak2024}. Yet other works propose eliminating basecalling entirely, extracting insights relevant for certain tasks directly from raw data~\cite{wan2022}.

While these works address some challenges facing modern genomics, each leaves aspects unresolved. Works utilizing traditional accelerators (e.g. GPUs) improve throughput yet fail to address data movement overhead, and require energy-hungry, non-portable, and expensive hardware. Similarly, energy-inefficient FPGA proposals have been applied only on older algorithms that lack DNN basecallers' improved accuracies while falling severely short of the throughput requirements for real-time basecalling~\cite{benton2022}. Finally, "read until" and basecall-free approaches have limited applicability; in contrast, real-time basecalling accelerates every genome analysis pipeline. Thus, this work seeks to address the challenges brought about by large sequencing data, by tightly coupling real-time basecalling with sequencing on the same edge device.

\hll{To this end, we introduce a Compute-in-Memory Basecalling Accelerator (\emph{CiMBA}), the most energy and area efficient basecalling accelerator demonstrated thus yet, capable of real-time basecalling within a small area/power envelope. CiMBA processes raw sequencing data via a deeply-pipelined dataflow architecture, while flexibly supporting a wide range of basecalling DNNs. Through the use of Non-Volatile Memory (NVM) crossbar arrays, CiMBA achieves SotA energy and area efficiency by eliminating all DNN weight motion and enabling highly parallel (up to 262K per tile) MAC operations, enabling basecalling exceeding the rate of data generation. CiMBA uses a 2D mesh to efficiently transport data-vectors between NVM-arrays that store weight values and perform Vector-Matrix Multiplications (VMMs) and digital processing units for activation functions. With a capacity of 2.9M weights, CiMBA can easily handle a range of networks, including Dorado-Fast (0.47M weights), ONT's SotA lightweight basecaller. CiMBA is also equipped with a LookAround Decoder, implementing a novel hardware/latency-conservative decoding strategy that enables high throughput. Sufficient DNN accuracy can be maintained despite analog induced noise through careful preparation~\cite{rasch2023} and programming~\cite{mackin2019, mackin2022} of DNN weights.}

CiMBA provides four significant advantages over existing works: (1) greatly reduced power (1.17W)  and (2) area ($25$mm$^2$) requirements, (3) \textgreater40$\times$ less device-to-workstation communication overhead, and (4) real-time, on-chip DNN-based basecalling. As a result, this accelerator enables a wide variety of applications, ranging from partial basecalling (using "read-until")~\cite{cavlak2024} to full basecalling, including metagenomics~\cite{2022shahroodi}, genomics~\cite{lou2020}, and the incorporation of downstream analysis accelerators such as KrakenOnMem~\cite{2022shahroodi}.

\begin{figure}
  \centering
  \includegraphics[width=0.95\columnwidth,trim={1cm 1cm 1cm 1cm},clip]{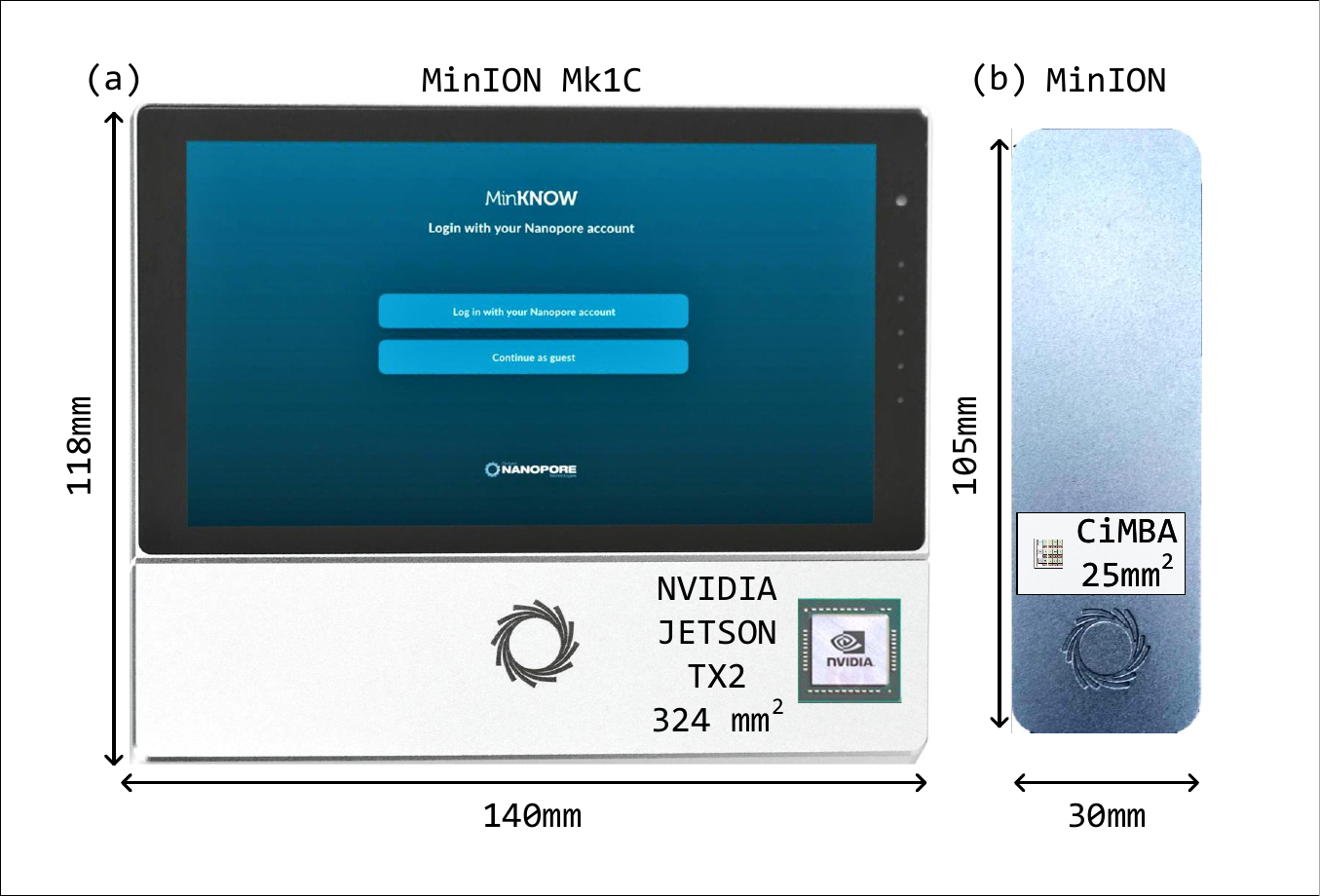}
  \caption{Size comparison of (a) the MinION Mk1C device that features MinION sequencing device and TX2 embedded GPU, and (b) the standalone MinION device along with our proposed CiMBA basecalling processor.}
  \label{fig:minion}
\end{figure}

To demonstrate CiMBA's flexibility in DNN support and deploy its full computational power, we also introduce AnaLog (AL)-Dorado, a new family of basecalling DNN models.
These models introduce optimizations that enhance basecalling performance while maintaining SotA accuracy on analog devices-- including hardware-verified analysis of layer sensitivity to CiM noise sources to mitigate accuracy loss, layer-size optimization for maximum crossbar array utilization and efficiency, and careful design of data-transport to minimize contention.
AL-Dorado is trained in a hardware-aware fashion, recouping accuracy lost to digital quantization and analog noise sources. 

Thus, the contributions of this work are as follows:
\begin{itemize}
\itemsep-0.1em 
    \item We propose CiMBA, a \underline{C}ompute-\underline{i}n-\underline{M}emory \underline{B}asecalling \underline{A}rchitecture, comprising of multiple NVM arrays for in-memory computation coupled to a novel LookAround Decoder, enabling real-time, on-chip, accurate transduction of incoming raw data into nucleotide sequences.
    \item CiMBA's mesh-based architecture supports a wide range of DNNs at high accuracy with a 8$\times$/13$\times$ area/power reduction against SotA embedded GPU accelerators.
    \item To complement CiMBA, we introduce the AL-Dorado line of DNN basecallers. By optimizing DNNs for realistic NVM devices, we mitigate the effects of analog noise on accuracy and maximize hardware utilization.
    \item AL-Dorado on CiMBA achieves 2$\times$/17$\times$/27$\times$ better throughput/energy consumption/compute density compared to Dorado-Fast on the Jetson Xavier AGX while maintaining 91\% basecalling accuracy.
\end{itemize}

\section{Background}
\label{sec:background}

Oxford Nanopore Technology's (ONT) SotA sequencing flow for generating ultra-long reads (up to 2.2 million bases) enables analysis of human genome regions inaccessible by other sequencing technologies~\cite{jain2018}, and recovery of highly-contiguous, even nearly complete, microbial genomes~\cite{sereika2022}. ONT's portable, handheld sequencing machines, the MinION and MinION Mk1C, illustrated in Figure~\ref{fig:minion}, enable diverse biomedical applications ranging from clinical diagnostics~\cite{ashley2016,flores2013} to environmental monitoring~\cite{deVries2022,cruz2023}.

ONT sequencing uses flow cells composed of nanoscale channel arrays containing nanopores. When a DNA molecule passes through the nanopore, an electrical current amplitude is disrupted over time, producing amplitude distortions known as "squiggles." Such raw electrical current signals are generally not analyzed directly since these data are perturbed by various noise sources such as variations in the movement speed of the DNA molecule through the channel. Thus, the transduction of raw data into the corresponding DNA sequence requires a complex and computationally expensive algorithm known as "basecalling." Algorithms incorporating DNNs can offer (at least) 10\% higher accuracy for nucleotide base prediction~\cite{wick2019}.

\subsection{Basecalling pipeline}
\label{subsec:basecallPipeline}
\begin{figure}
  \centering
  \includegraphics[width=0.85\columnwidth,trim={1cm 1cm 1cm 1cm},clip]{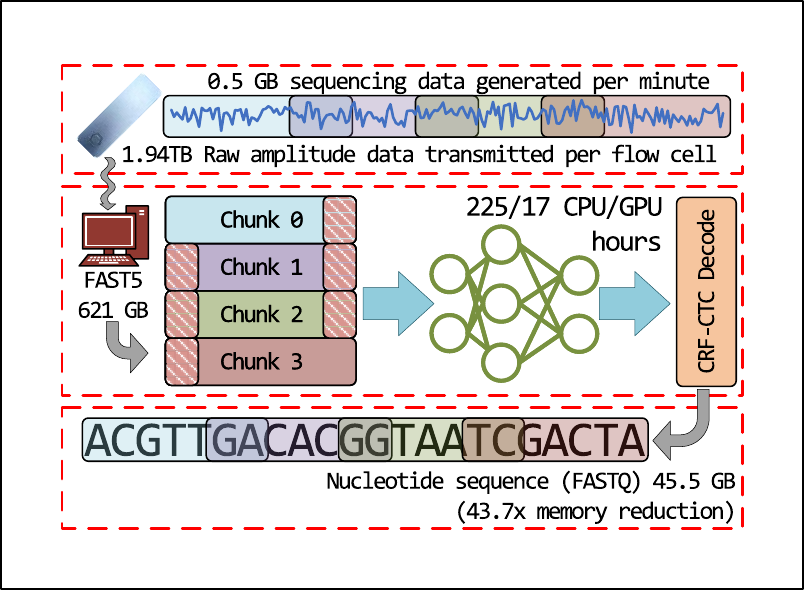}
  \caption{
  The MinION produces $\sim$0.5 GB of raw signal data/minute to be streamed to workstation for basecalling, which then incurs 40\% (NVIDIA Xavier AGX) to 86\% (Xeon W-10885M) of the sequencing pipeline due to large parameter counts/DRAM access costs common to LSTM DNNs~\cite{9563028}.}
  \label{fig:basecallpipline}
\end{figure}
Figure~\ref{fig:basecallpipline} illustrates the basecalling pipeline, spanning data generation, splitting the data into chunks, inferring chunk base sequences via DNN, and finally reassembly into long-reads.\\

\begin{figure}
  \centering
  \includegraphics[width=0.9\columnwidth,trim={1cm 0.98cm 1cm 1cm},clip]{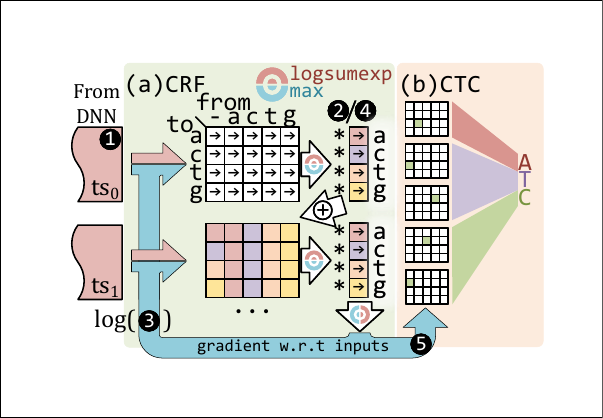}
  \caption{CRF-CTC decoding (state length=1 for simplicity): DNN outputs represent log-likelihoods of state-transitions (\ding{182}). Path likelihood is accumulated across timesteps (\ding{183}/\ding{185}), and gradient of the final value w.r.t. the inputs (\ding{184}) is iteratively evaluated to identify each timestep's most likely transition (\ding{186}).}
  \label{fig:crf-ctc}
\end{figure}

\begin{spacing}{1.01}
\textbf{Raw signals from flow cells --} 
Current amplitudes are sensed as DNA/RNA strands pass through a flow cell's nanopores. A flow cell contains up to 512 channels each capable of simultaneous sequencing at a sampling frequency of 4kHz, resulting in a maximum data generation rate of 0.46GB per minute, or 1.94TB data for the duration of the flow cell's $\sim$72 hour lifetime. As a frame of reference,~\cite{Gorzynski2022} utilized $\sim$3.46 flow cells, generating 5.8TB of usable data, to sequence a single human genome. This data explosion is set to continue as sequencing technology continues to improve; for example, the larger PromethION flow cell is capable of generating \textgreater100x the base pairs of the MinION~\cite{flowcell}.\\ %
\textbf{Data splitting/stitching --} 
Since these raw data cannot be basecalled directly, they are instead split into chunks and grouped into batches for basecalling. These chunks are typically overlapped with previous and subsequent chunks, and re-stitched into long-reads after inference. The default values of chunk size of 4000 and overlap of 500 provided by the Bonito framework~\cite{oxford2023} causes 25\% of bases to be basecalled twice, leading to extra computation.\\
\textbf{DNN basecalling --} 
After data splitting, a DNN basecaller model is used to infer the nucleotide sequence. There are many to choose from:
SotA models consist of hybrid networks using Convolutional (CNN) layers as feature extractors and Long Short-Term Memory (LSTM) layers to learn the temporal relationship between timesteps. Currently, ONT recommends its closed-source Guppy network~\cite{wick2019}. Also under active development are the research-oriented Bonito~\cite{wright2020} and Dorado~\cite{accesswire2022} model families, the latter comprising of 3 networks of varying size and performance: \textit{Fast} (0.59 GigaMACs per inference), \textit{High Accuracy} (5.15 GMACs), and \textit{Super Accuracy} (21.6 GMACs). Even the relatively small Dorado Fast model requires a high-end, energy-/area-intensive embedded GPU to achieve real-time basecalling, albeit at lower accuracy~\cite{benton2022}. We will consider Dorado Fast as a baseline for comparison in this work, demonstrating how the CiM paradigm alleviates these bottlenecks by co-locating storage and computation on-tile.\\
\textbf{CRF-CTC decoding --} 
Historically, basecalling networks have used Connectionist Temporal Classification (CTC) decoders, as illustrated by Figure~\ref{fig:crf-ctc}-b. With CTC decoding, the probability of each base is predicted by the DNN at each timestep. The final nucleotide sequence is predicted by collapsing repeating bases into a single base, with a learned blank space dividing consecutive repeating bases. While CTC decoding thus allows inference to be agnostic to sampling frequency, it assumes conditional independence between timesteps, which fails to capture the physical reality of the nanopore signal. 

To this end, modern basecallers add a Conditional Random Field (CRF) step to the decoder, to capture conditional dependence between timesteps~\cite{pages2023}. Rather than predicting the base directly, the basecaller predicts transitions between states, where a state represents a sequence of bases. To calculate the Transition Probability (TP) at each timestep, the DNN outputs (\ding{182} in Figure~\ref{fig:crf-ctc}-a) are first arranged such that each row (column) represents all transitions to (from) a state. The likelihood of a transition $p(b \rightarrow b')$ at a given timestep is defined by 1) the transition's TP at that timestep (as given by the DNN), 2) the probability of arriving at state $b$ via previous timesteps (the sum of all TP's ending in $b$ at the previous timestep (\ding{183})), and 3) the probability of being in state $b'$ at the next timestep (calculated by taking the gradient of a final scaler value w.r.t the DNN output after all timesteps have been processed (\ding{184})). In order to calculate the Max Likely Path (MLP), \ding{183}/\ding{184} are repeated, replacing the summation with a max function~(\ding{185}/\ding{186}). The final output is the most likely transition at each timestep, accounting for all transitions at all timesteps. While predicting transitions instead of states using a gradient significantly boosts accuracy~\cite{pages2023}, this incurs additional memory and compute costs, since all timesteps (by default 800 amplitude values) must be inferred and stored before the gradient can be calculated. For this reason, while CRF-CTC decoding is used to train basecalling DNNs, traditional CTC methods are still used in the Bonito and Dorado frameworks. We address this computational and memory overhead in Section~\ref{subsec:lookaround}.
\end{spacing}

\subsection{Compute-in-Memory (CiM) paradigm}
\label{subsec:cim}
\label{subsec:crf}

\ronenew{Computation near or even using memory elements and circuitry is a non-von Neumann paradigm that comes in a variety of flavors, of which each is suitable to accelerate the computation of different workloads and algorithms. As an example in the bioinformatics domain, the utilization of digital}~\cite{lanius2024}, \ronenew{analog}~\cite{zhang2023}, \ronenew{and Content Addressable Memory (CAM) architectures}~\cite{harary2024,kaplan2020} \ronenew{are being widely explored for aligning read sequences to reference genomes with low latency and extremely high energy efficiency, due to their ability to perform operations at the point of data storage. Similar benefits at the basecalling step can be achieved via the Compute-in-Memory (CiM) paradigm utilizing crossbar arrays}~\cite{Y2020sebastianNatNano}. 
These systems' amenability for multiply-and-accumulate (MAC) operations make the CiM paradigm popular for DNN acceleration, since MACs comprise $>$98\% of the operations in widely-used DNN benchmarks~\cite{Y2023jainTVLSI}. Previous works have studied the use of CiM for the basecalling step~\cite{lou2018, lou2020, mao2022, shahroodi2023swordfish}, most notably Helix~\cite{lou2020}. This work's relation with previous CiM works is discussed in Section~\ref{sec:related}.

SRAM is a widely-studied technology for building CiM arrays owing to technological maturity and scalability~\cite{Y2022jiaJSSC}. In contrast, non-volatile memory (NVM) CiM offers increased area density and the elimination of weight-transport available only with full weight-stationarity~\cite{Y2023jainTVLSI}. More traditional NOR-Flash~\cite{Y2022fickISSCC}, 2D NAND-Flash~\cite{Y2018merrikhbayatIEEETran}, and 3D NAND-Flash~\cite{Y2022kimJSSC}, as well as emerging NVMs (eNVMs) including Resistive Random Access Memory (RRAM)~\cite{Y2022wanNature, Y2021HungNatElec}, Magnetoresistive Random Access Memory (MRAM)~\cite{Y2022jungNature} and Phase-Change Memory (PCM)~\cite{LEGALLO202063} are actively explored as memory primitives for CiM acceleration. We consider PCM in this work due to its being arguably the most mature eNVM technology~\cite{Salahuddin2018, stpcm}, with high-capacity analog storage (up to 4 bits per synaptic cell~\cite{Y2023legalloNatElec}), and superior endurance~\cite{Salahuddin2018}. However, the CiMBA accelerator architecture is ultimately NVM technology agnostic and can be adopted for other technologies.  

\begin{figure}
  \centering
  \includegraphics[width=\columnwidth,trim={1cm 1cm 1cm 1cm},clip]{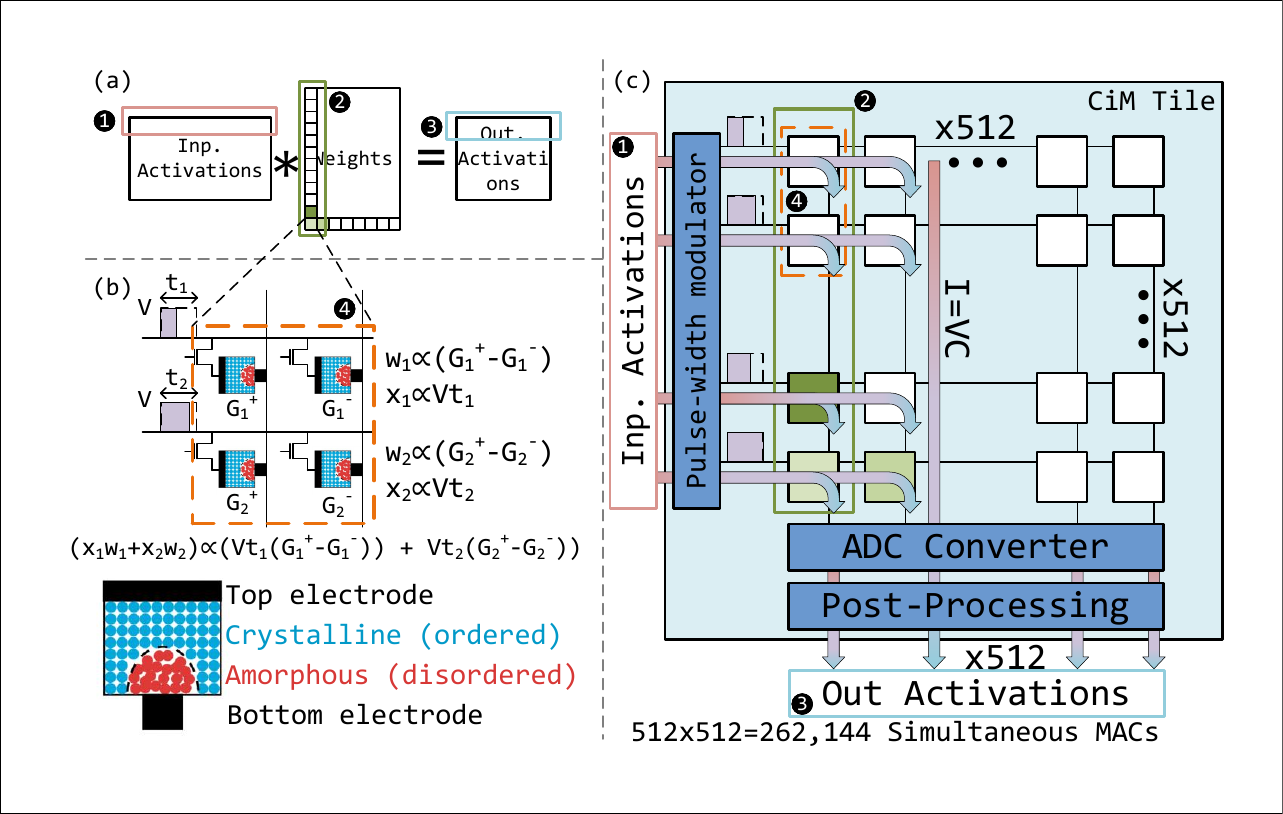}
  \caption{(a) Mapping a DNN layer to (b) an array of eNVM cells (c) enables massively parallel MAC operations.}
  \label{fig:AIMC}
\end{figure}

A PCM CiM tile architecture is shown in Figure \ref{fig:AIMC}-a. Each synaptic unit-cell typically includes two PCM devices, to accommodate signed weights. Each PCM cell consists of a phase-change material between two electrodes, with values programmed by adjusting the PCM material between a high-conductance crystalline and low-conductance amorphous structure via joule heating~\cite{LEGALLO202063}. Unlike conventional memory arrays, numerous rows of a CiM array are enabled simultaneously. The resulting current along the bitline represents the dot product between an input vector (introduced onto the rows by pulse width modulation) and a weight vector (encoded into the conductances of a column of unit cells). 

\hll{We note that previous works have explored the utility of CiM in the basecalling domain~\cite{lou2020, shahroodi2023swordfish}. Sections~\ref{sec:cimba} and \ref{sec:al-dorado} describe CiMBA's hardware and algorithmic uniqueness in relation to these works, while Section~\ref{sec:related} directly assesses the three works in relation to each other.}

\subsection{\hll{Enabling efficient, parallel DNN operations on CiM tiles}}
\label{subsec:tile}
Figure~\ref{fig:AIMC}-b/c shows how the weights of a matrix may be mapped to a CiM tile. Mapping fully-connected layers is performed in such a manner, while convolutional and LSTM layers may be implemented on CiM arrays in a similar fashion. For convolutional layers of kernel geometry $c_{in} \times k_w \times k_h \times c_{out}$, kernels are converted to $c_{out}$ columns of height $c_{in} \times k_w \times k_h$ before being mapped onto CiM tiles, allowing output channels to be calculated in parallel. For LSTM layers, the set of weights are mapped to the CiM array in an interleaved fashion to minimize routing for the subsequent auxiliary operations~\cite{Y2023legalloNatElec}. 

Properly mapping weights into tiles enables highly parallel computation of the billions of MAC operations required by Dorado networks described in Section~\ref{subsec:basecallPipeline}. Indeed, fully filling a 512x512 CiM tile results in 262K simultaneous MAC operations per tile. With an integration time of 40ns~\cite{Y2023jainTVLSI}, this enables a performance of 6.55 TOPS per tile, scaling by the number of tiles on-device. Such tile-level performance can only lead to high {\em system}-level performance if interconnect can maintain a comparable throughput, motivating a massively parallel 2D mesh as described in Section~\ref{subsec:2Dmesh}.

Further, by performing computations where the data resides in a highly parallel manner, CiM arrays offer increased energy efficiency over conventional approaches. 10 TOPS/W is reported for recent PCM CiM hardware successfully integrated in 14nm CMOS node through back-end-of-the-line processing~\cite{Y2023legalloNatElec}, with next generation designs expected to exceed this~\cite{Y2021burrIEEESpectrum}. The results presented in Section~\ref{sec:results} demonstrate how analog CiM greatly accelerates basecalling at extremely low power and area overheads.

\section{Challenges \& Motivation}
\begin{scriptsize}
\begin{table}[t]
    \caption{Communication/storage overhead for 9 datasets~\cite{wick2019_2} is reduced by 43.7$\times$/4.37$\times$ via on-chip basecalling.}
    \centering
    \resizebox{\columnwidth}{!}{
    \begin{tabular}{r | c |c c | c c c}
                  &     & \multicolumn{2}{c|}{Communication} & \multicolumn{3}{c}{Storage (GB)}    \\
                   &    & \textbf{Raw}          & \textbf{Nucleotide}        &            &            &             \\
    \textbf{Dataset}         & \textbf{Reads}   & \textbf{Sequence (GB)}       & \textbf{String (GB)}      & \textbf{FAST5}      & \textbf{POD5}       & \textbf{FASTQ}       \\\hline\hline
    Acinetobacter    & 4,467 & 4.80         & 0.11        & 1.5        & 0.97       & 0.35        \\\hline
    Haemophilus & 8,669       & 5.79         & 0.07        & 1.8        & 1.2        & 0.36        \\\hline
    Klebsiella INF032   & 15,154    & 18.86        & 0.52        & 6.1        & 4.1        & 1.5         \\\hline
    Klebsiella INF042    & 11,278   & 22.53        & 0.51        & 7.0        & 4.6        & 1.7         \\\hline
    Klebsiella KSB2   & 15,178     & 16.76        & 0.38        & 5.3        & 3.5        & 1.3         \\\hline
    Klebsiella NUH29   & 11,047    & 12.25        & 0.23        & 3.9        & 2.5        & 0.844       \\\hline
    Serratia      & 	16,847     & 5.59         & 0.13        & 2.0        & 1.3        & .44         \\\hline
    Staphylococcus     & 	16,742    & 9.04  & 0.23 & 2.9 & 1.9 & 0.68 \\\hline
    Stenotrophomonas  & 16,010      & 22.60 & 0.46 & 7.2 & 4.7 & 1.6  \\\hline
    \textbf{Total}  & 115,392   & 118.6        & 2.7         & 37.6       & 24.77      & 8.6         \\\hline
    \textbf{Reduction}& & --           & \bf{43.7x}  & --         & 1.5x       & \bf{4.37x}  \\\hline
        \end{tabular}}
    \label{tab:datastorage}
\end{table}
\end{scriptsize}

\label{sec:challenges}
Basecalling faces a number of throughput, communication and storage challenges that we seek to address in this work. Further, our proposed CiM solution introduces its own challenges which are discussed in the coming sections.

\subsection{Real-time genome analysis}

Portable sequencing enabled a wide range of applications beyond clinical scenarios~\cite{borsting2015,maria2017,cruz2023}, driven in large part by ONT's introduction of the MinION Mk1C, a sequencing device with an onboard Jetson TX2 embedded GPU~\cite{nvidiajetson}, pictured in Figure~\ref{fig:minion}-a. However, even at the current level of flow cell technology, the TX2 has trouble maintaining real-time throughput~\cite{benton2022}, and further improvements to flow cell technology will soon require more compute power than is offered by the embedded GPU~\cite{dunn2021,oxford2022}. Further, as mentioned in Section~\ref{subsec:basecallPipeline}, CRF-CTC decoding with gradients requires chunks of data to be inferred completely before decoding can occur. Such a "pipeline bubble" greatly complicates implementation of an efficient,
end-to-end sequencing pipeline. This work thus proposes an embedded analog basecalling accelerator that can address both of these challenges.
\subsection{Data communication and storage}
\begin{figure}
  \centering
  \includegraphics[width=\columnwidth,trim={1cm 1cm 1cm 1cm},clip]{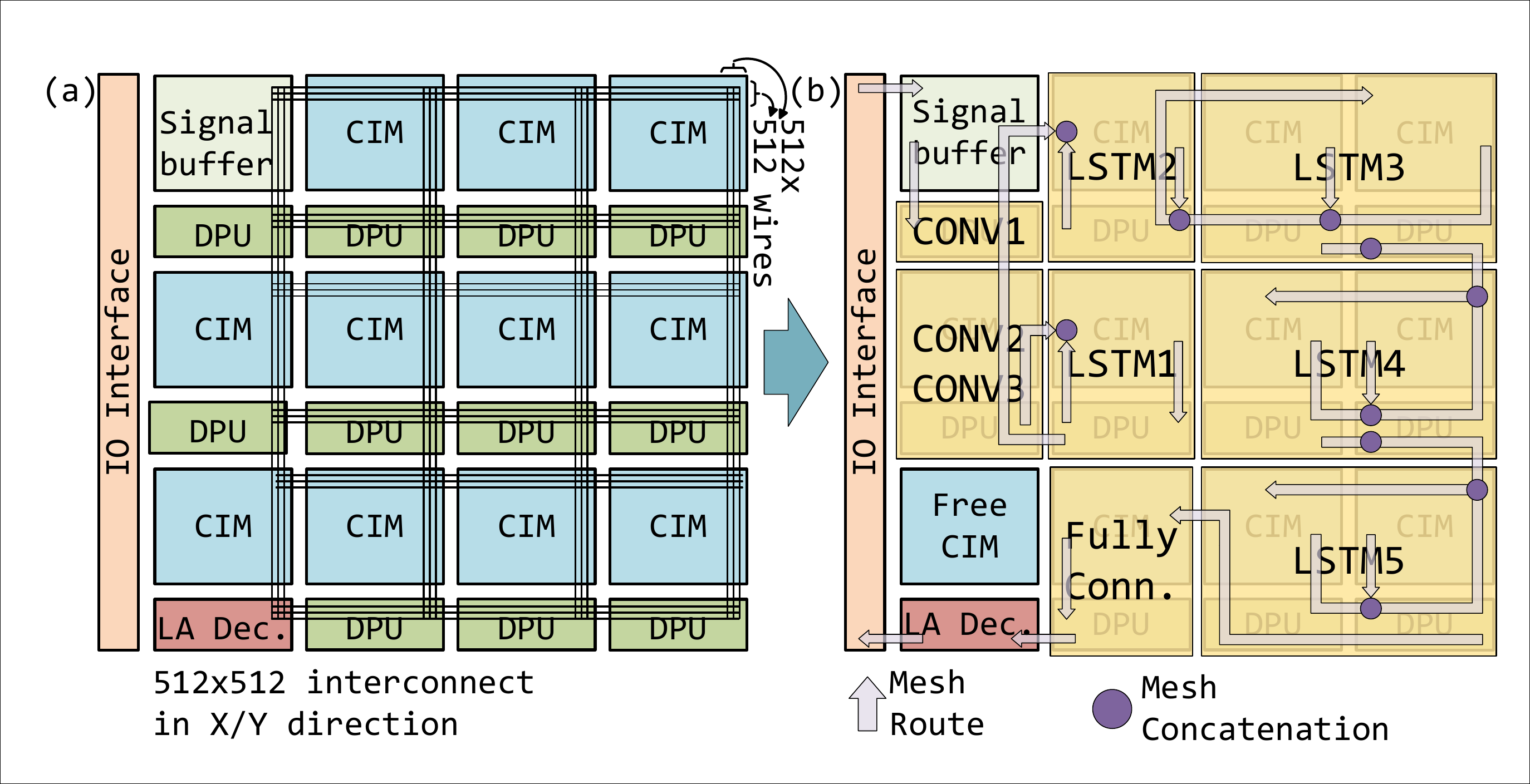}
  \caption{Mapping of AL-Dorado (Figure~\ref{fig:dorado}) on the CiMBA architecture. }
  \label{fig:cimba}
\end{figure}
The challenge of real-time basecalling arises from the vast amount of data generated during sequencing. Table~\ref{tab:datastorage} illustrates the dichotomy between raw signal data and final nucleotide sequence size. At a sampling frequency of 4kHz, roughly 10 floating point values are generated per base. By performing basecalling on-device, this raw data can be converted to 8-bit base values \textit{before} transmission, reducing communication (storage) overhead by 43.7$\times$ (4.37$\times$). While the Mk1C can support current flow cell technology, future sequencing devices are expected to further increase data-rate and 
-volume, calling for enhanced basecalling accelerators that can continue to implement on-device basecalling. 

\subsection{CiM noise sources}
\label{subsec:noise}

A key drawback of CiM approaches is the reduced precision arising from various noise sources\hll{, most critically those creating discrepancies between the desired and actual stored weight.} Some of this error occurs when programming the synaptic weights onto the conductance values of the NVM devices. These conductance values change over time due to the intrinsic structural relaxation of the amorphous phase (conductance drift)~\cite{Y2021boybatIEDM}, and due to read noise~\cite{Y2020JoshiNatComms, Y2020nandakumarIEDM}. Other sources of imprecision can arise from the peripheral circuitry, or from quantization noise associated with data conversion. Although these noise sources can induce considerable performance loss if DNN models are deployed naively onto analog hardware~\cite{Y2020JoshiNatComms}, noise-aware offline training~\cite{Y2020JoshiNatComms, rasch2023}, weight-to-conductance mapping~\cite{mackin2019, mackin2022}, and iterative weight-programming methods~\cite{Y2021narayananITED} can overcome these challenges. There are also device-level innovations such as projected PCM that could improve the compute precision substantially \cite{Y2018giannopoulosIEDM}.

\subsection{CiM-amenable model architectures}
\label{subsec:CiM-model}
Not all network architectures are amenable for analog acceleration. For instance, depth-wise-based bottleneck layers~\cite{sandler2019mobilenetv2} results in poor array utilization (<0.1\%)\cite{Y2022garofaloJETCAS}, addressed in~\cite{Y2022zhouIEEEMicro} by alterations in the model architecture or in~\cite{Y2022garofaloJETCAS} by a hybrid digital-analog acceleration. \hll{Furthermore, layers with uneven row/column aspect ratios or tiny kernels} may result in under-utilization of the CiM arrays~\cite{Y2023jainTVLSI}. Model-architecture co-design is therefore critical to fully benefit from CiM acceleration. 

\section{CiMBA architecture}
\label{sec:cimba}

To meet the aforementioned data generation\?communication bottlenecks and overcome the unique design challenges presented by CiM noise, we propose a Compute-in-Memory Basecalling Accelerator architecture (CiMBA), illustrated in Figure~\ref{fig:cimba}. 
CiMBA is a 25mm$^2$ module which supports a wide range of DNN basecallers at extremely low power and with high throughput. It is composed of mixed-signal CiM tiles (Figure~\ref{fig:AIMC}-b), custom Digital Processing Units (DPUs) (Figure~\ref{fig:dpu}), a LookAround (LA) decoder (Figure~\ref{fig:chainAddArch}), and a signal buffer (Figure~\ref{fig:cimba}). These heterogeneous blocks communicate through a 2D mesh-based interconnect (Figure~\ref{fig:cimba}) \hll{with a regular structure that enables flexible mapping of DNN layers, enabling optimal data-flow pipelining, while also facilitating future scaling of the architecture to support larger models.}

\begin{scriptsize}
\begin{table}
    \caption{CiMBA supported operations}
    \centering
    \resizebox{0.85\columnwidth}{!}{
    \begin{tabular}{l l}
    \hline\hline
    \textbf{Component} & \textbf{Supported operations} \\\hline\hline
    \mr[2]{CiM tile} & Analog VMM, \\
    & post-processing (MUL/ADD)\\\hline
    \mr[3]{DPU}       & Digital VMM, affine scale (MUL/ADD),  \\ 
    & BatchNorm (MUL/ADD), LUT, \\
    & Activation alignment (memory)\\\hline
    LA Dec.   & LookAround Decoding  \\\hline
    Signal buffer & Memory \\\hline\hline
        \end{tabular}}
    \label{tab:architecture-ops}
\end{table}
\end{scriptsize}
\subsection{CiM tiles}

CiMBA's 11 CiM tiles feature PCM-based crossbar arrays comprising of a 512$\times$512 synaptic unit cells, 512 Analog-to-Digital Converters (ADCs), 512 Pulse Width Modulators (PWMs) and a small digital post-processing block. \hll{The CiM tile data flow is shown in Figure~\ref{fig:AIMC}-c.} In contrast to previous SotA works ISAAC~\cite{Y2016shafieeISCA} and Helix~\cite{lou2020}, which extends ISAAC, CiMBA employs a larger 512$\times512$ crossbar to a) enable mapping of the first three LSTM layers of Dorado Fast and AL-Dorado to a single tile, reducing mapping and routing complexity, and b) to amortize tile periphery overhead over a larger unit cell count. The technological feasibility of a crossbar with 512$\times$512 unit-cells has been demonstrated~\cite{Y2021narayananITED}, and larger arrays are predicted for future CiM tiles~\cite{Y2021burrIEEESpectrum}. PWMs provide 8-bit signed input data for the crossbar, while each bitline in CiMBA is connected to a compact Current Controlled Oscillator (CCO)-based ADC~\cite{Y2022-khaddam-aljameh-JSSC}. The resulting currents are digitized and accumulated by the CCO-based ADCs to produce a 10-bit signed integer. The digital post-processing block helps adjust for ADC gain variations caused by circuit-level mismatch.

CiMBA's 11 CiM tiles can achieve a theoretical maximum of 72 TOPS, or 2.88 TOPS/mm$^2$. In comparison, the Xavier AGX embedded GPU presented in Section~\ref{subsec:baselines} achieves a peak TOPS/mm$^2$ of 0.57. We would therefore expect CiMBA to significantly outperform the Xavier in experimental results.

\subsection{2D mesh based architecture} \label{subsec:2Dmesh}

As specified in Section~\ref{subsec:tile}, a high bandwidth interconnect is needed to support the CiM tiles' low latency and energy efficiency on the system-level. Moreover, the interconnect should allow flexibility on mapping workloads to the heterogeneous fabric, as neural network model architectures may undergo rapid changes. For CiMBA, we adopt a 2D mesh as described in~\cite{Y2023jainTVLSI} and illustrated in Figure~\ref{fig:cimba}-a, to \hll{move activations between nodes.} This mesh comprises of multiple sets of parallel wires, running in the X and Y direction and crossing over each node. The 2D structure allows independent data-transfers along X and Y directions, thereby minimizing the distance between any two components, while maintaining several parallel data-transfers. The mesh is capable of implicit concatenation of vectors from multiple sources (e.g. input and hidden vectors destined for LSTM layers) and multi-casting to multiple destinations. Further, given the determinism of the networks in question, standard handshake protocols between nodes can be foregone, drastically improving energy efficiency while further reducing area overhead. 

Figure~\ref{fig:cimba}-b illustrates the highly parallel nature enabled by the 2D-mesh's high throughput, deeply pipelined implementation. Unless otherwise stated, digital computations and mesh transfers are performed in INT10, chosen for its non-negligible performance gain over INT8 in analog CiM tasks~\cite{Spoon2021}.

\subsection{Digital Processing Unit (DPU)}
\label{subsec:digital}

\begin{figure}
  \centering
  \includegraphics[width=0.9\columnwidth,trim={1cm 1cm 1cm 1cm},clip]{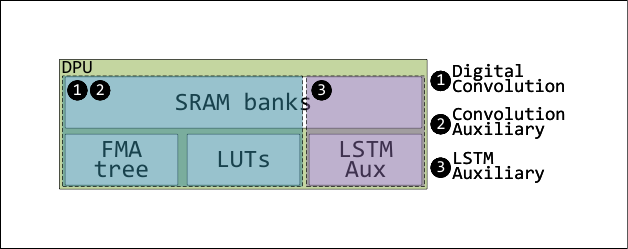}
  \caption{Overview of the DPU structure.}
  \label{fig:dpu}
\end{figure}

CiMBA's DPU blocks, presented in Figure \ref{fig:dpu}, support a variety of computational flows, including (1) digital convolutions, (2) auxiliary operations such as activation functions or batch normalization, and (3) LSTM digital auxiliary operations, all with 16-bit floating point precision. SRAM banks in each DPU support the memory requirement of these flows. 

\begin{figure*}
  \centering
  \includegraphics[width=\textwidth,trim={1cm 1cm 1cm 1cm},clip]{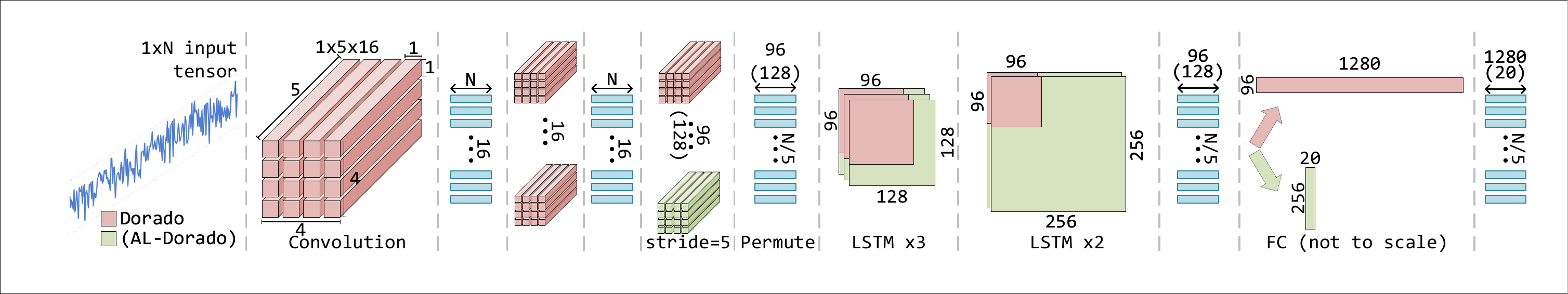}
  \caption{Dorado (red) and AL-Dorado (red/green) DNN architectures.}
  \label{fig:dorado}
\end{figure*}
\textbf{Digital convolutions -} First, weights, bias, and activations are read from the SRAM banks. If the activations arrive externally to the unit, they are first scaled to 16-bit floating point format. Next, convolution is performed via a tree of Fused Multiply-and-Add (FMA) units. Swish and clamp are executed via a Look-Up Table (LUT)~\cite{Y2022-khaddam-aljameh-JSSC}, followed by an FMA which applies the selected piecewise-linear slope and offset. Lastly, the 16-bit floating point result is converted to a 10-bit integer format for transfer via mesh. 

\textbf{Convolution auxiliary -} VMM results arrive to the DPU from a CiM tile. They are scaled from a 10-bit integer format into a 16-bit floating point format. Next, the batch normalization parameters are loaded from the SRAM banks and batch normalization is executed via FMAs. Any additional affine scaling\cite{rasch2023} is \ implemented via this one set of scaling parameters. This is followed by the swish activation, implemented using a LUT and FMA. Clamp operations are implicitly supported by setting the upper/lower bounds of the LUT transfer function. Finally, the output is converted to 10-bit integer format. 
A similar flow handles digital operations following fully connected layers executed on CiM tiles.

\textbf{LSTM auxiliary -} The VMM result is converted from a 10-bit integer into 16-bit floating point format. This is followed by the application of an affine scaling operation with the FMAs. The bias of the LSTM weights are included in the additive factor. Next, the LSTM auxiliary operations are performed, which include element-wise ADD, element-wise MUL, tanh/sigmoid activation functions. ADD and MUL blocks within the DPU handle the element-wise operations, while the LUTs, followed by FMAs perform the tanh/sigmoid activations. The interleaved mapping ensures that the previous cell state is stored in a small SRAM, replaced by the new cell state as appropriate. The output is converted from a 16-bit floating point representation into a 10-bit integer. 

\subsection{Lookaround decoder}

In order to address the decoding challenges presented in Section~\ref{subsec:basecallPipeline}, we developed a LookAround decoder block to accelerate the decoding step. This block's functionality will be discussed in detail in Section~\ref{subsec:lookaround}.

\subsection{Signal buffer}
The MinION flow cell is capable of generating data on 512 channels simultaneously, with each channel sequencing a distinct nucleotide strand. Hence, it is essential to buffer raw signals from each channel as they are captured, and process them individually. CiMBA therefore incorporates a signal buffer for this purpose. The signal buffer is an SRAM-rich component with a memory controller to orchestrate the data flow (1) from the IO interface to the SRAM, and (2) from the SRAM to the DPU for the convolution operation. Each of the 512 channels is allocated 2.45kB of memory, and more than 1000 raw read signals can be stored per channel. The overall SRAM capacity in the signal buffer is 1.25MB. 

\section{AL-Dorado Models for CiMBA}
\label{sec:al-dorado}
As described in Section~\ref{subsec:noise}, DNN inference on CiM arrays is highly sensitive to analog noise and prone to resource under-utilization. 
We developed a set of networks addressing these constraints, and we detail the design choices of one such \underline{A}na\underline{L}og (AL)-Dorado network in this section.

\subsection{Dorado Fast GPU implementation baseline}

As the CiMBA architecture is to be implemented near-flow cell and be capable of real-time basecalling, we take as a baseline the Dorado series of DNNs~\cite{oxford2023_2}. The Dorado series are hybrid CNN-LSTM networks consisting of 3 1D CNN layers followed by 5 LSTM layers and 1 Fully-Connected (FC) layer. The output values of the FC layer represent transition log-probabilities as described in Section~\ref{subsec:basecallPipeline}. Dorado comes in three flavours, \textit{Fast}, \textit{High Accuracy} and \textit{Super Accuracy}. We use Dorado Fast as the base model for our AL-Dorado networks, as it consists of only 0.47 million weights, making it amenable to embedded CiM acceleration. Dorado Fast is also specifically designed to "keep up" with Nanopore's sequencing devices, while still providing high accuracy, and therefore represents the current SotA in real-time basecalling DNN models. Dorado Fast is illustrated in Figure~\ref{fig:dorado}. 

\subsection{AL-Dorado model architecture}
\label{subsec:aldorado}
Upon the Dorado Fast architecture, we explored a range of architectural modifications to optimize the network for CiMBA. The AL-Dorado networks were developed through several design/experiment iterations, which will be elaborated on in Section~\ref{sec:results}. For brevity, only one studied AL-Dorado network is presented here, illustrated in Figure~\ref{fig:dorado}. Specifically, the LSTM size is boosted from dimensions of 96 to 128 for layers 1-3, and to 256 for layers 4-5 to account for the heterogeneous layer response to analog conversion detailed in Section~\ref{subsec:hetero}. We reintroduce the clamp layers between the CNN layers and after the FC layer present in the higher accuracy Dorado models, as these can be handled implicitly by CiMBA's LUT tables within the DPUs and provide an accuracy gain. Finally, we reduce the possible output state lengths to 1, resulting in an output of 20 transition probabilities per timestep, for the reasons outlined in~\ref{subsec:decodingresults}. These modifications result in a network consisting of 1.7M weight parameters, placing the network midway between Dorado Fast (0.47M) and HAC (6.2M).

While we present only one AL-Dorado model in this work, the range of potential networks is large. Other network-amenable modifications are under exploration, including reducing the initial layers' noise sensitivity via larger kernels or layer collapsing~\cite{bhardwaj2022}, increasing output layer state length while accommodating for LookAround decoding, including extra linear layers as in the Dorado HAC and SUP models and mapped to the unused CiM tile in Figure~\ref{fig:cimba}-b, or increasing the size of individual LSTM layers as enabled by CiM tile size constraints. Further additions to the line of AL-Dorado networks will be detailed in future works.

\subsection{LookAround decoding}
\label{subsec:lookaround}
\begin{figure}
  \centering
  \includegraphics[width=\columnwidth,trim={1cm 1cm 1cm 1cm},clip]{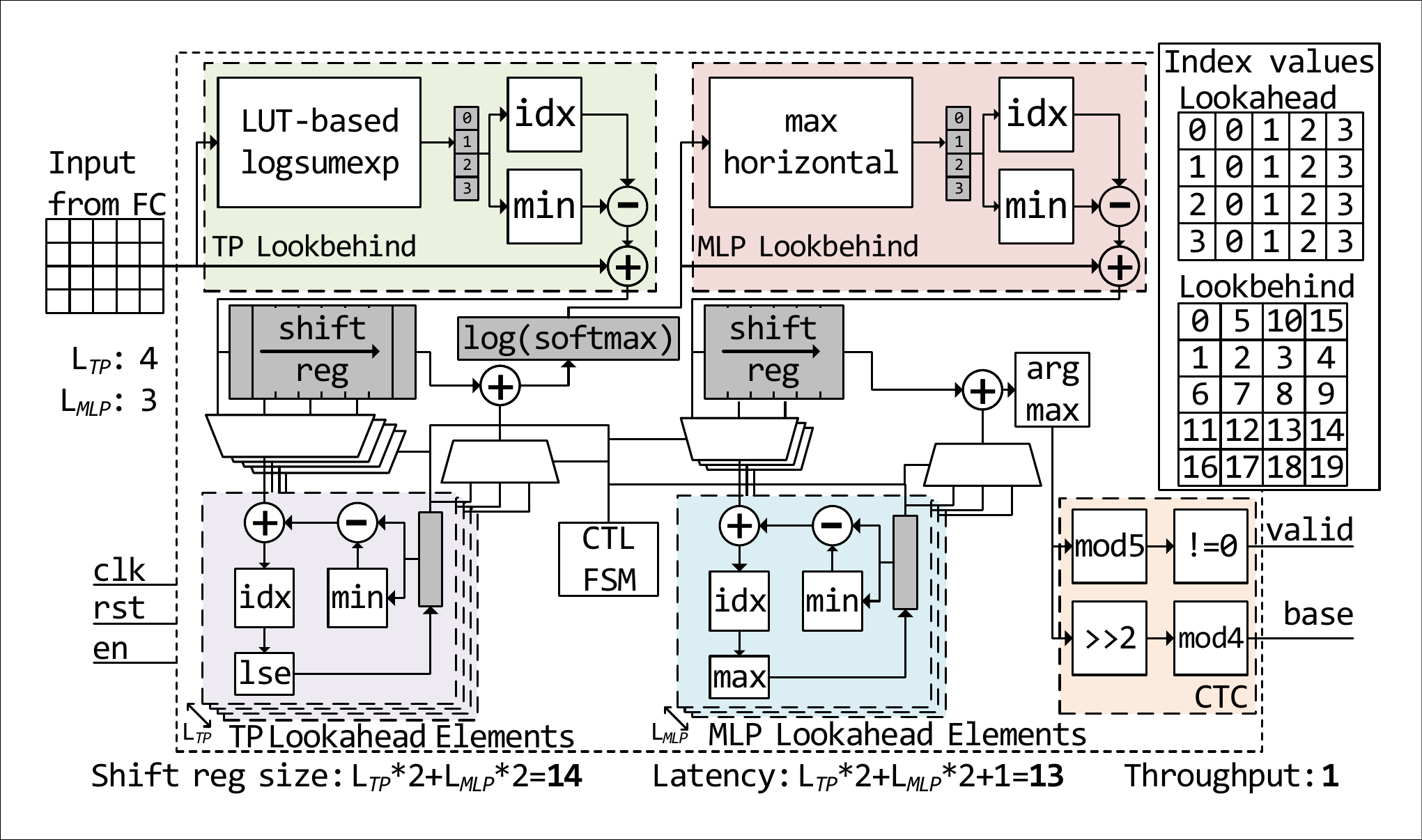}
  \caption{Architecture of the LookAround decoder.}
  \label{fig:chainAddArch}
\end{figure}
As described in Section~\ref{subsec:basecallPipeline}, CRF-CTC using gradients requires all timesteps to be computed before decoding the nucleotide sequence, rendering it not amenable for a streaming basecalling architecture. We therefore propose another style of decoding we name LookAround (LA) decoding. Instead of considering all timesteps for computing the probability and most likely state of each timestep, the LA decoder considers only the transitions in a given window. Namely, for timestep $T_n$, the timesteps considered are $T_{n-1}$ (Lookbehind) and $T_{n+L}$ (Lookahead), where $L$ is a tunable parameter for both the TP and MLP portions of the decoding process. As $L_{TP}$ or $L_{MLP}$ increases, more future timesteps are considered, thus asymptotically approaching CRF-CTC w/gradient accuracy. This enables a tradeoff between accuracy and area/latency, as will be discussed in Section~\ref{subsec:decodingresults}. Figure~\ref{fig:chainAddArch} illustrates the block diagram of the LA decoder. It is divided into symmetrical halves, calculating first the TP values, then the MLP values. Unlike gradient CRF-CTC decoding, timesteps are discarded once no longer needed, greatly reducing memory requirements and improving throughput. Namely, $2*L_{TP}+2*L_{MLP}$ registers are necessary to hold the requisite number of timesteps, and a latency of $2*L_{TP}+2*L_{MLP}+1$ cycles is incurred for decoding. It should be noted that the number of parallel Lookahead elements is equal to the values of $L$ for each half, thus maintaining a throughput of 1 sample processed per cycle.

\section{Methodology}
\label{sec:methodology}

To comprehensively explore CiMBA/AL-Dorado's feasibility as a real-time basecaller, we analyze both CiMBA's runtime basecalling characteristics, as well as study the impact of CiM noise on AL-Dorado in terms of single inference as well as downstream analysis accuracy.

\subsection{SotA comparison methodology and frameworks}
\label{subsec:baselines}
We compare CiMBA's performance against a range of devices used for basecalling. Firstly, as a benchmark to demonstrate server level basecalling performance, we benchmark Dorado v0.3.3 on an NVIDIA A100 by basecalling 20k reads stored in POD5 format. Dorado is ONT's most advanced basecalling framework that is highly optimized for the A100/H100 GPUs~\cite{oxford2023_2} with performant optimizations such as INT8 weight quantization. Indeed, Dorado has been recently shown to outperform ONT's proprietary Guppy basecaller by 1.96x~\cite{ads2023}. We also benchmark against the TX2 and Xavier AGX embedded GPUs via results sourced from~\cite{benton2022}. As these results were performed using the Guppy framework with FAST5 inputs, we scale them by 3.2x, the ratio between our A100 results and those found in~\cite{benton2022}. We note that this scaling factor is generous, given that embedded GPUs are unable to take full advantage of Dorado's optimizations.

We also compare against the SotA Helix~\cite{lou2020} and DeepCoral~\cite{perešíni2021} edge devices for low-power, low-area basecalling. Helix is an extension of the ISAAC architecture specialized for basecalling, while DeepCoral accelerates basecalling with the Google Coral Edge TPU DNN accelerator~\cite{Herzog2022}.  Comparing HW/SW co-designed works such as these is challenging, due to the impossibility of isolating either the architecture or algorithm under study. We note, however, that each work strives to be a low-power, embedded basecalling accelerator, and on this basis we believe some useful comparison on orders of magnitude can be made. For Helix, we report results from the Guppy network with 0.244M weights.

\subsection{CiMBA architecture system-level simulation environment}

To verify AL-Dorado system level performance on the CiMBA architecture, we perform system level simulations using the simulation tool described in the work by Jain~\textit{et al.}~\cite{Y2023jainTVLSI} This simulator enables highly parameterizable, cycle accurate simulation of 2D mesh-based CiM architectures. It accepts as input an architecture definition, a network description, and mapping of the network on the architecture. It then generates a highly granular list of interdependent micro-operations that capture all aspects of the network graph, including VMM and digital operations, mesh transfers, and memory accesses. This job graph is scheduled on the system architecture to capture mesh and resource contention, dependency stalls, routing energy and performance overhead, etc. Table~\ref{tab:architecture-params} presents key but non-exhaustive architectural parameters defining the CiMBA architecture in the simulation tool, and Figure~\ref{fig:cimba}-b illustrates the mapping of AL-Dorado onto CiMBA. All blocks are synthesized in Cadence Genus and physically implemented in Cadence Innovus in 14nm FinFET technology and clocked at a 1GHz frequency \rtwonew{to verify their functionality for future fabrication efforts.}

\begin{scriptsize}
\begin{table}
    \caption{Accelerator architecture parameters}
    \setlength{\tabcolsep}{2pt}
    \centering
    \resizebox{0.85\columnwidth}{!}{
    \begin{tabular}{l l l l}
    \hline\hline
    & \textbf{Operation}/ & \textbf{Value/} & \textbf{Latency} \\
    \textbf{Component} & \textbf{Parameter} & \textbf{Energy} & \textbf{(cycles)} \\\hline\hline
    
    \mr[5]{CiM tile}    & VMM, 512x512 unit cells               & 5.2nJ & 40  \\
                        & \rtwonew{Max cell conductance}                  & 25{\textmu}S & \\
                        & \rtwonew{Read noise std. dev}                   & 0.1 & \\
                        & \rtwonew{Programming noise std. dev}            & 1.0 & \\\hline
    \mr[4]{DPU}         & BatchNorm, ADD, MUL                   & 1.24pJ  & 3 \\
                        & LUT, Swish                            & 1.49pJ  & 4 \\
                        & LSTM auxiliary                        & 19.3pJ  & 25\\
                        & SRAM R/W per bit         & 2.5fJ      & 1 \\\hline
    \mr[3]{2D mesh}     & East-West per bit        & 44.9 fJ    & 3 \\
                        & North-South per bit      & 81.4 fJ    & 3\\
                        & Turn per bit             & 126 fJ     & 3\\\hline
     \mr[3]{LA decoder} & $L_{TP}$                 & 4          &  \\
                        & $L_{MLP}$                & 1          &  \\
                        & Decode                   & 0.16nJ     & 11 \\\hline
    Signal buffer       & SRAM R/W per bit         & 2.5fJ      & 1\\\hline
        \end{tabular}}
    \label{tab:architecture-params}
\end{table}
\end{scriptsize}

\subsection{AL-Dorado training HW/SW environment}
Training is performed using A100 GPUs via distributed data parallel training. We use the Bonito software repository as a base for developing the AL-Dorado model~\cite{oxford2023} rather than the newer Dorado repository as its Python implementation enables more expedient development/experiment iterations, however, the latest Dorado-Fast model is ported from the Dorado repository. The network is trained until validation accuracy saturates, a total of 30 epochs.

\hll{To study the impact of analog noise sources described in Section~\ref{subsec:noise}} and develop the mitigation strategies detailed in Section~\ref{subsec:aldorado}, we use the AIHWKIT~\cite{rasch2023} Python library. AIHWKIT enables training and inference in an analog-aware manner that takes into account analog non-idealities such as the dynamic range of input voltages, weights, and outputs, noise injected by weight quantization and programming noise, DAC/ADC quantization, and effects of weight changes due to PCM conductance drift over time. AIHWKIT accepts as input a digital network as well as a tile configuration defining PCM parameters and tile width and height. The tile configuration is generalized to the flavour of NV memory under study and the resultant network can be mapped on any device utilizing the same technology. The AL-Dorado network is trained for 29 epochs in floating point, then converted to analog in AIHWKIT and retrained for a further 5 epochs.

\subsection{PCM hardware validation}

\begin{figure}
  \centering
  \includegraphics[width=\columnwidth,trim={0cm 0.7cm 0cm 0.7cm},clip]{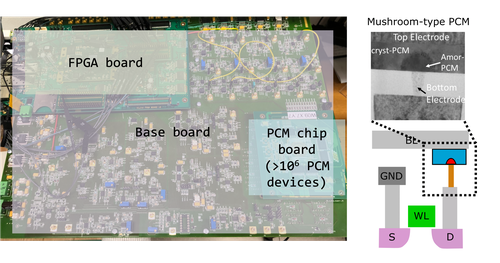}
  \caption{\rtwonew{Hardware platform utilized to characterize read, write, and drift characteristics of PCM memory array for simulation of the CiMBA platform.}}
  \label{fig:fusion}
\end{figure}

In order to define AIHWKIT's tile configuration, we also have at our disposal a physical PCM memory array consisting of \textgreater1 million PCM cells allowing single device read/writes, pictured in Figure~\ref{fig:fusion}~\cite{Y2010closeIEDM}. \rtwonew{We characterize the read, write, and drift characteristics of this chip and configure AIHWKIT to simulate these characteristics. Table}~\ref{tab:architecture-params} \rtwonew{indicates array key characterization parameters. The exact functionality of AIHWKIT is beyond the scope of this work, and we recommend the reader to}~\cite{buchel2024} \rtwonew{for a detailed description of the implemented methodologies. It suffices to say that we verify} that AIHWKIT emits expected accuracies by programming our analog-aware trained weights into the PCM array and measuring their drift over one day. Inference with these drifted values is verified to match AIHWKIT's predicted accuracies.

It should be noted that the PCM array characteristics of this hardware platform (e.g., drift behavior, conductance ranges, unit cell design) as well as their models within AIHWKIT may be slightly different from that of a presumed CiMBA prototype, which represents an architecture using next generation PCM devices. This is owed to a variety of factors, including PCM geometry and process technology. It is not expected therefore that AL-Dorado implemented on a physically fabricated CiMBA prototype would behave identically to the results presented here; however, the trends shown will be consistent across hardware.

\label{subsec:fusion}

\subsection{Training/validation/bacterial datasets and limitations}
Training and validation DNA datasets collected on R9.4.1 flow cells are downloaded from the Bonito repository~\cite{oxford2023}. They consist of 65k/1000 full reads for training/validation, respectively, split into chunks of 4000. The validation dataset is used to analyze single chunk accuracy of Dorado-Fast and AL-Dorado in floating point and on CiMBA, as each validation chunk comes with a reference sequence.

To verify accuracy beyond inference accuracy, we also perform post-basecalling analysis on a set of reads generated using a MinION R9.4.1 flowcell. Table~\ref{tab:datastorage} provides details on the 9 evaluated organisms~\cite{wick2019_2}. We note that these publicly available datasets use an older chemistry and lack certain accuracy-boosting features present in newer datasets, such as the Duplex read method~\cite{duplex}, capable of boosting raw read accuracy to \textgreater99\%. Therefore, experimental results show a level of accuracy that would be expected from such datasets, and we expect that our method would benefit platforms using newer chemistry as well.

\subsection{Post-basecalling analysis flow}

We evaluate AL-Dorado performance using aligned basecalling accuracy, i.e., the total number of exactly matched bases between a read and the reference genome divided by the total alignment length including insertions and deletions. 

We basecall each read set, producing either a FASTQ or FASTA file suitable for downstream analysis. We align each basecalled read to its corresponding reference genome of the same species using the state-of-the-art read mapper, minimap2~\cite{li_minimap2_2018}. We use Rebaler~\cite{rebaler} to generate a consensus sequence from each basecalled read set before polishing the genome with multiple rounds of Racon~\cite{vaser2017fast}. This approach ensures that the assembled genome will have the same large-scale structure as the reference.

\section{Results}
\label{sec:results}
This section details analysis results on both the CiMBA architecture and the AL-Dorado DNN basecaller. We analyze performance both in terms of the throughput and power consumption of CiMBA, as well as the accuracy implications of performing basecalling on a future CiMBA prototype.
\begin{figure}
  \centering
  \includegraphics[width=\columnwidth,trim={1cm 1cm 1cm 1cm},clip]{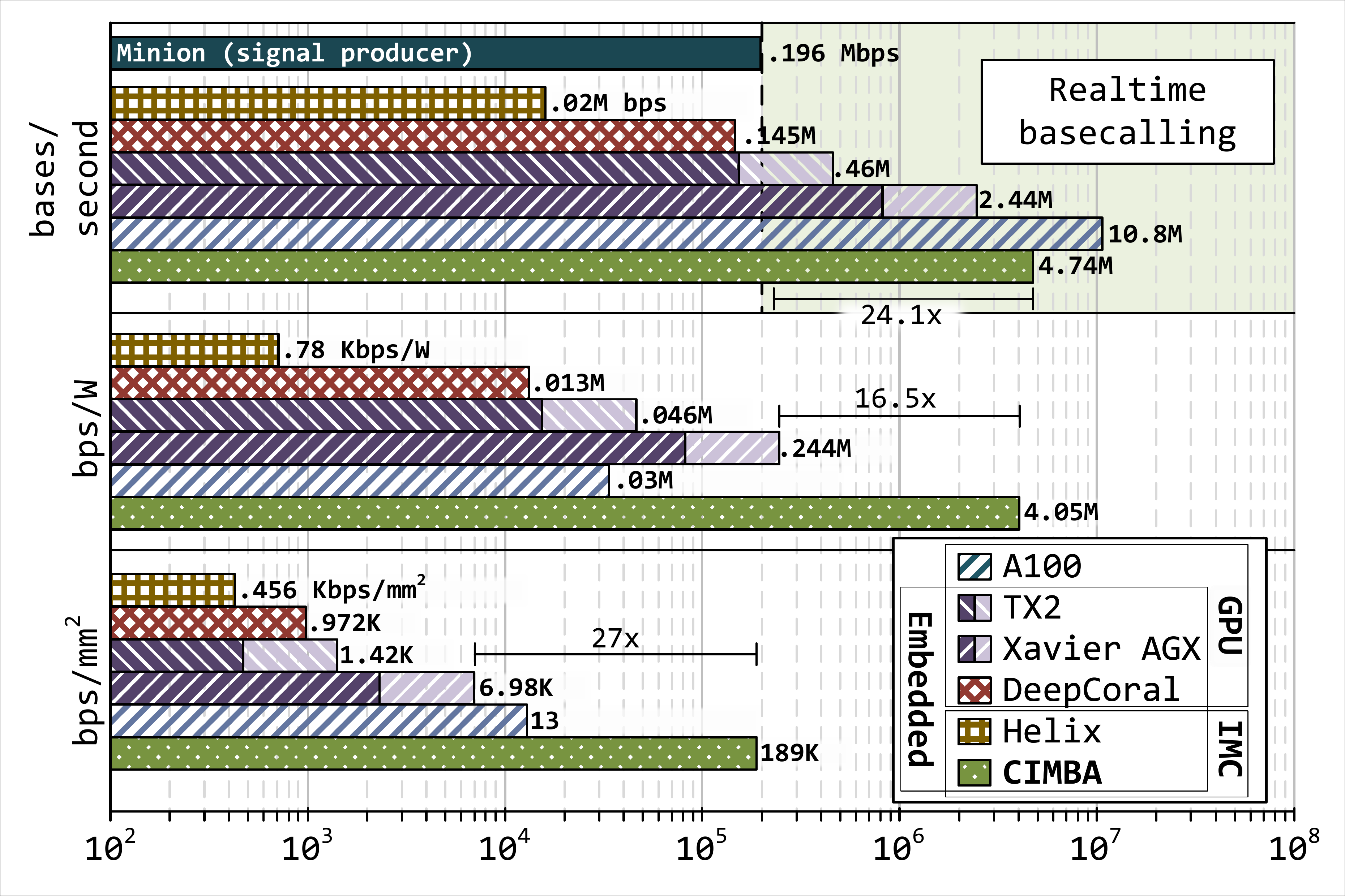}
  \caption{Comparison between basecalling architectures. TX2 and Xavier extensions represent scaling to account for Dorado optimizations. CiMBA significantly outperforms all architectures in terms of bps/W and bps/mm$^2$ and achieves 24x the throughput necessary to operate in real-time.}
  \label{fig:results}
\end{figure}
\subsection{CiMBA performance analysis}
\label{subsec:perf}

Figure~\ref{fig:results} illustrates CiMBA's performance against the SotA baselines described in Section~\ref{subsec:baselines}. \hll{As can be seen, CiMBA's throughput outperforms} all devices except the A100, as expected when comparing against a data-center level GPU. However, when throughput is balanced against power and area footprint, CiMBA outperforms all other embedded devices by at least 16.5$\times$/27$\times$ in terms of bps/W and bps/mm$^2$, respectively. At 25mm$^2$ and with an average power consumption of 1.17W, \hll{CiMBA favorably compares to the MinION Mk1C's 322mm$^2$ embedded GPU in terms of power and area efficiency. Section~\ref{sec:related} provides analyis of performance improvement over Helix.}

\subsection{CiMBA runtime analysis}
\label{subsec:runtimeanalysis}
\begin{figure}
  \centering
  \includegraphics[width=\columnwidth,trim={1cm 1cm 1cm 1cm},clip]{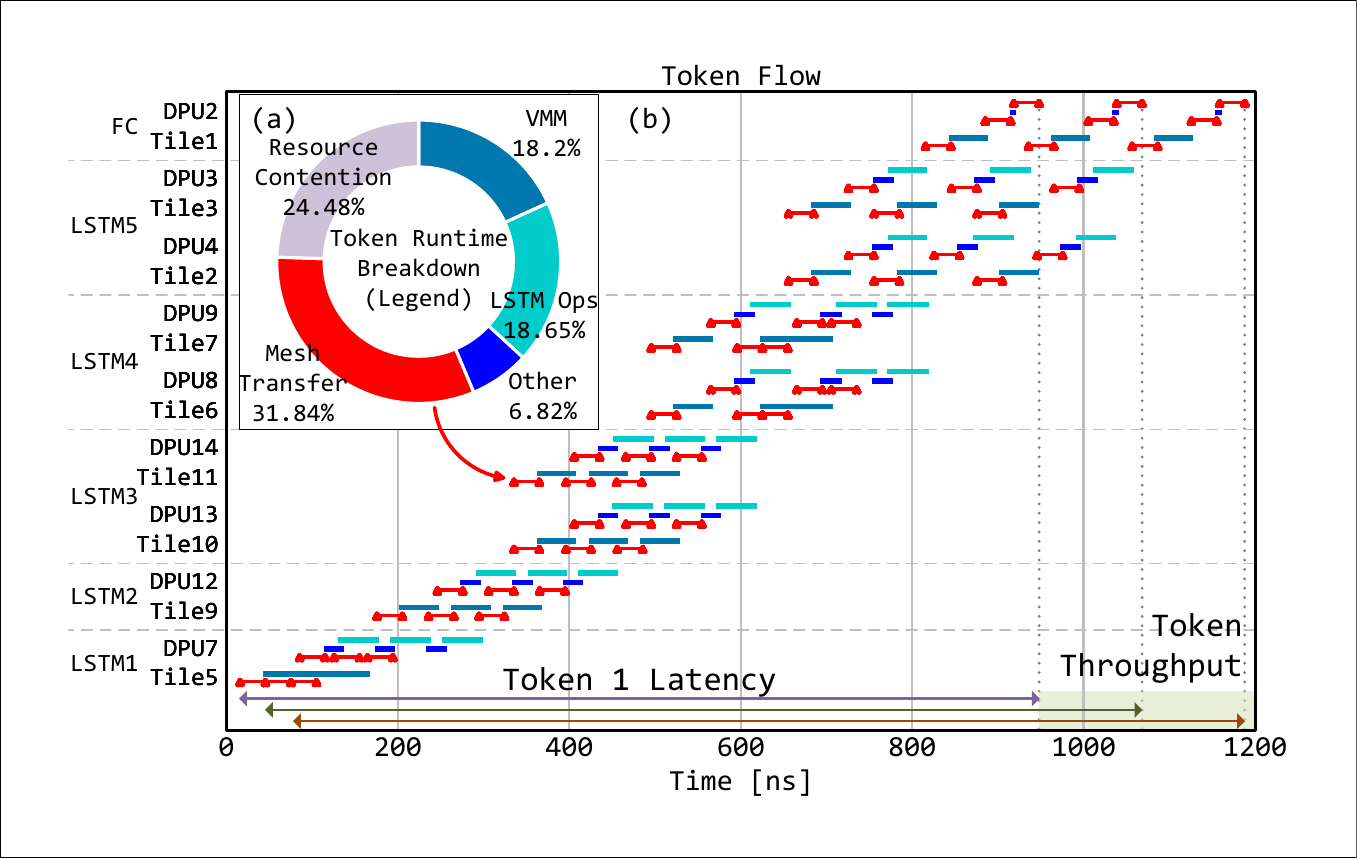}
  \caption{(a) Runtime profile for a token. Resource contention limits performance, visible as white gaps (stalls) between the colored lines (active computation) in (b), which tracks three tokens as they move through the network. Despite this, (b) illustrates how high token throughput is achieved via pipelining in CiMBA.}
  \label{fig:ads}
  \vspace{-10pt}
\end{figure}

Figure~\ref{fig:ads} highlights a subset of interesting results gained from our system level simulations of CiMBA. Specifically, we are able to see in (a) the breakdown of runtime into different op categories. VMMs refer to the LSTM matrix multiplication operations, while LSTM Ops account for all operations required to calculate the input, forget, cell, and output gates, and the hidden state. Other includes CNN, clamp, batchnorm, and LA decoder operations. Resource contention accounts for any time an operation must wait for a preceding operation to release a resource, which primarily occurs when multiple mesh transfers require the same portion of the mesh. 

Taken together, it can be seen that data movement accounts for roughly 60\% of the total runtime. This indicates that network mapping on CiMBA is critical. Mapping layers in relation to their data sources and destinations played an important role in improving network throughput.

\hll{The benefits of CiMBA's data pipelining strategy is illustrated in Figure~\ref{fig:ads}-b,} which illustrates the movement of 3 tokens through the LSTM portion of the network. Tokens are processed in parallel on different nodes within the mesh before being sent to the next destination. Intra-layer parallel computation also occurs in layers that are too big to fit in one CiM tile, as is apparent on tiles 10 and 11 in Figure~\ref{fig:ads}-b, which jointly compute LSTM layer 3. This pipelining enables CiMBA to achieve extremely high throughput as described in Section~\ref{subsec:perf}, achieving 24x the required bases-per-second necessary to perform real-time basecalling.

\subsection{\hll{Accuracy implications of analog conversion and retraining}}
\begin{figure}
  \centering
  \includegraphics[width=\columnwidth,trim={1cm 1cm 1cm 1cm},clip]{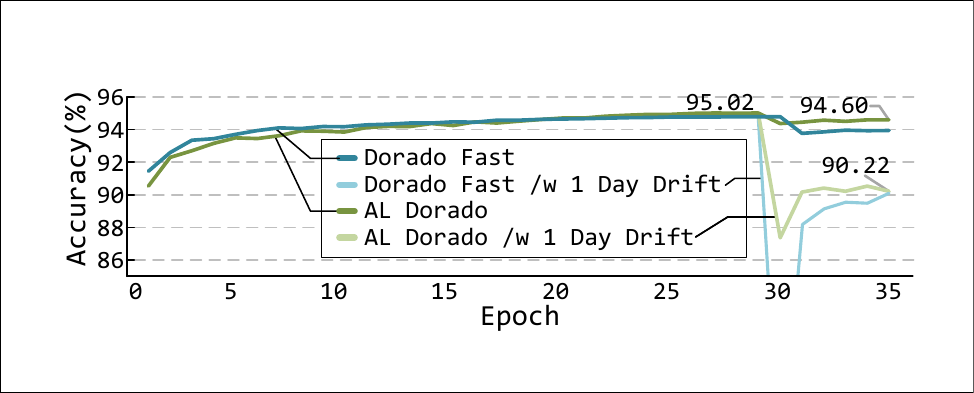}
  \caption{Networks are digitally trained for 29 epochs, then converted to analog and CiM-aware trained for 5 epochs. While this conversion incurs little accuracy loss, CiM noise due to drift greatly impact accuracy, even after retraining.}
  \label{fig:training}
  \vspace{-5pt}
\end{figure}

Figure~\ref{fig:training} reports the accuracy after FP training as well as \hll{pre-/post-} analog-aware retraining. Analog accuracies are recorded with and without a drift of one day. Standard CRF-CTC decoding is used to isolate the impact of drift. \hll{While analog conversion degrades accuracy, loss is} recovered with retraining, particularly for AL-Dorado due to its CiM amenable architecture. On the other hand, accuracy loss due to drift is significant for both networks, and must be further addressed.

\subsection{Model sensitivity to analog nonidealities}
\label{subsec:hetero}
\begin{figure}
  \centering
  \includegraphics[width=0.9\columnwidth,trim={1cm 1cm 1cm 1cm},clip]{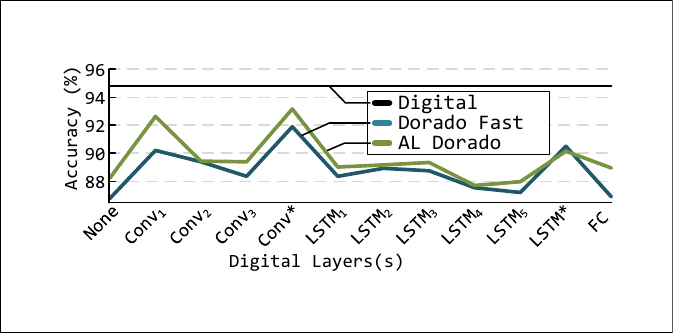}
  \caption{Layer sensitivity is analyzed by measuring network accuracy while maintaining each layer in digital. Results show that the first convolutional layer is highly sensitive to analog-induced noise.}
  \label{fig:hetero}
  \vspace{-10pt}
\end{figure}

To address PCM drift, we measure individual layers' sensitivity to analog noise by maintaining portions of the network in digital while converting the rest of the network to analog, illustrated in Figure~\ref{fig:hetero}. As can be seen, layers are not impacted equally by analog execution. Namely, it is clear that the CNN layers, particularly the first layer, are highly sensitive to analog noise. This can be explained by the layer's 1x5 kernel; as only 5 PCM cells contribute to the analog VMM, these cells are highly sensitive to noise. This observation aligns with work in other domains such as image recognition, where initial feature extraction layers are very susceptible to analog noise~\cite{Y2020JoshiNatComms}. 

Knowledge gained in this analysis motivates the design choice detailed in Section~\ref{subsec:digital}, namely, digitally computing the first layer. As layer 1 contains only 80 weight values, performing it in a digital node incurs no extra latency, as the CNN portion of the hybrid network has a higher throughput in relation to the LSTM portion. The benefits of maintaining the first layer in digital will be apparent in the next sections.

\subsection{Drift}
\label{subsec:drift}
\begin{figure}
  \centering
  \includegraphics[width=\columnwidth,trim={1cm 1cm 1cm 1cm},clip]{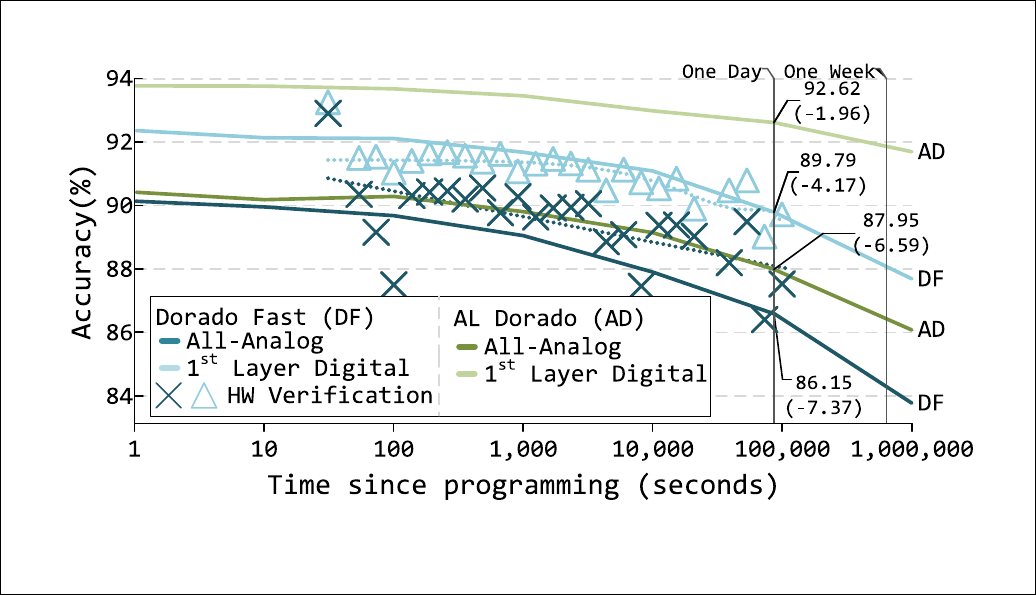}
  \caption{Impact of weight drift is apparent (-7\% accuracy for Dorado Fast). Digitally computing the first layer improves accuracy (-4.17\%), and AL-Dorado's analog optimizations further reduce lost accuracy (-1.96\%).}
  \label{fig:drift}
\end{figure}

Figure~\ref{fig:drift} illustrates the impact of PCM drift on network accuracy. Each line represents the accuracy of either Dorado Fast or AL-Dorado simulated by AIHWKIT over the course of roughly 11 days, while $\times$'s and $\triangle$'s represent inference using actual weights sampled over the course of a day on physical PCM hardware as described in Section~\ref{subsec:fusion}. As can be seen, accuracy loss due to drift is substantial; in the case of Dorado Fast, a \textgreater7\% accuracy drop is measured in both AIHWKIT and hardware. Digitally computing the first layer reduces this loss to 4.17\%, while the analog optimized AL-Dorado network incurs only a 1.96\% loss. While periodically refreshing the PCM tiles with the original analog-aware trained network weights remedies drift, proper network design and mapping mitigates accuracy loss in between reload periods.

\subsection{Decoding}
\label{subsec:decodingresults}

\begin{figure}
  \centering
  \includegraphics[width=0.8\columnwidth,trim={1cm 1cm 1cm 1cm},clip]{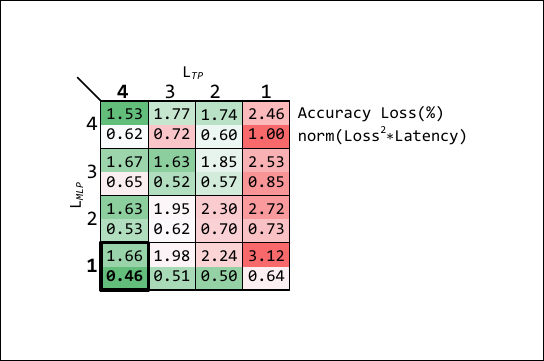}
  \caption{Grid analysis of look-ahead distances for LA decoding. At iso-latency a deeper $L_{TP}$ in contrast to $L_{MLP}$ yields higher accuracy.}
  \label{fig:decoding}
  \vspace{-5pt}
\end{figure}

As described in Section~\ref{subsec:basecallPipeline}, CRF-CTC decoding requires all timesteps to be inferred before decoding. In an effort to enable streaming basecalling, we introduced our LA decoder in Section~\ref{subsec:lookaround}. As this decoder only considers timesteps before and a limited number of timesteps after the step under consideration, accuracy is expected to drop in comparison to the baseline. To quantify this, we sweep values of $L_{TP}$ and $L_{MLP}$ between 1 and 4, meaning that we consider between 1 and 4 future timesteps for path likelihood and most likely path calculation, respectively. Figure~\ref{fig:decoding} illustrates the results. It can be seen in the upper values that accuracy is improved as either variable increases; however, when loss and latency/area overhead are considered jointly, increasing $L_{TP}$ has a greater benefit. As such, in Section~\ref{subsec:downstream}, we use AL-Dorado with LA values of 4 and 1 for $L_{TP}$ and $L_{MLP}$, respectively.

Currently, LA decoding is limited to a state size of 1, a limitation that is offset by gains in throughput/latency/power due to the smaller final layer and less complicated decoding block. Methods for extending the AL strategy to larger state sizes is ongoing and will lead to significant accuracy gains.

\subsection{Downstream analysis}
\label{subsec:downstream}

\begin{figure}
  \centering
  \includegraphics[width=\columnwidth,trim={1cm 1cm 1cm 1cm},clip]{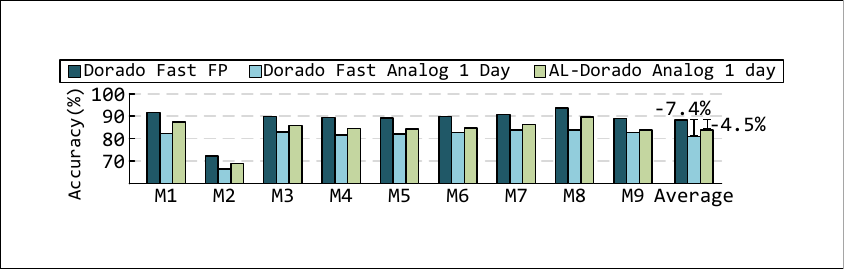}
  \caption{Downstream analysis of Dorado Fast in FP and in analog after 1 day of drift, and AL-Dorado, over the 9 microbial datasets specified in Table~\ref{tab:datastorage}.}
  \label{fig:downstream}
\end{figure}

Figure~\ref{fig:downstream} illustrates the downstream analysis of the 9 microbial datasets listed in Table~\ref{tab:datastorage}. The network's varied performance across datasets aligns with previous research on the same dataset~\cite{mao2022, Rezaei2023}. We observe that the accuracy loss between Dorado Fast in floating point and its analog equivalent, along with the loss for AL-Dorado with the aforementioned LA decoder parameters, is consistent with the values reported in Sections~\ref{subsec:drift} and~\ref{subsec:decodingresults}. This demonstrates that AL-Dorado's CiM aware retraining and optimization strategies generalize beyond the training/validation sets used to implement them. 

Discussion of downstream analysis leads to a wider discussion of the general applicability of the concepts presented in this work, namely, the 'sequence-and-forget' method of on-device sequencing and discarding raw data to reduce communication and storage overhead, as well as acceptable accuracy for different use-cases. In a hospital or research setting where copious storage and compute power are accessible, it is beneficial to store raw sequence data for basecalling on future models, thus gaining accuracy. There are otherwise few analyses performed directly on raw sequence data, as outlined in Section~\ref{sec:related}-C, with downstream analysis such as taxonomic classification and variant calling performed on sequenced bases. Within these analyses, those falling into the field of metagenomics benefit most from CiMBA, as they require less basecalling accuracy to achieve meaningful results. Enabling in-field metagenomics opens application opportunities ranging from classifying a patient's saliva profile at a routine checkup~\cite{Wood2014} to DNA barcoding of endangered species in remote environments~\cite{Menegon2017}, in which MinION reads with error rates as 17\% were sufficient for achieving inter-species differentiation. We also note, without obviating the aforementioned use-cases at CiMBA's current levels of accuracy, that analog-based inference accuracy is an active research topic, improving as training and noise mitigation strategies are developed~\cite{Lammie2024, Khwa2021}.

\section{Related Work}
\label{sec:related}

\hll{To our knowledge, CiMBA is the \emph{first} embedded in-memory accelerator capable of performing real-time basecalling.} %
\hll{In this section, we describe other related works categorized in the following domains.}

\textit{A. CiM basecalling acceleration --} In comparison to previous CiM based works~\cite{lou2018, lou2020, mao2022, shahroodi2023swordfish}, specifically Helix~\cite{lou2020}, CiMBA and AL-Dorado are able to operate in real-time at a 16.8x reduced power requirement. Two differences that explain the drastic performance improvement of CiMBA over Helix are that Helix relies on energy and latency intensive writes to NVM during inference, and utilizes smaller arrays that are underutilized, e.g. in some cases only the diagonal is considered. Further, while little information is provided on model mapping and data routing for Helix, it may well be that CiMBA's 2D mesh more efficiently serves its 11 tiles to maintain peak tile throughput. Moreover, Helix focuses on quantized basecalling, which for instance simplifies analog weight storage to discretizing values, while the analog noise modeling adopted in this work provides a more realistic representation observed in CiM prototypes~\cite{Y2023legalloNatElec}.

Other related works include Swordfish~\cite{shahroodi2023swordfish}, which proposes using large a DNN ($\sim$27M weights) and accelerator (\textgreater270mm$^2$), falling outside the target embedded range of CiMBA, as well as not reporting power analysis.
CiMBA can be directly exploited for improving GenPip~\cite{mao2022}, an in-memory basecalling/read mapping accelerator, as a drop-in replacement for its Helix basecalling module. KrakenOnMem~\cite{2022shahroodi}, an in-memory taxonomic profiler, also presents interesting integration opportunities in terms of extending CiMBA's functionality. \rtwonew{We also note that CiMBA operates in the weight stationary domain, mitigating concerns about PCM endurance that commonly impact such works. Even considering a periodic refresh of the weight values, cell endurance far outlasts other reasons for device failure.}

\textit{B. Non-CiM basecalling acceleration --} \hll{Modern basecalling algorithms are based on DNNs amenable to GPU acceleration. Many works~\cite{wick2019, singh2024, goenka2022, xu2021,lv2020, zeng2020, yeh2022} accelerate modern and prior basecallers using GPUs.} FPGA acceleration~\cite{wu2019, hammad2021} for basecalling has also gained attention in the last few years. Recent works also exploited specialized processors, such as the Coral accelerator featuring the Edge Tensor Processing Unit~\cite{perešíni2021}, \hll{and the AMD-Xilinx Versal AI Engine~\cite{singh2024}.}

Unlike this work, existing non-CiM basecalling accelerators 1) do not fundamentally address the severe data movement overhead, and 2) usually require powerful GPUs with large chip area and high energy consumption to operate at reasonable speeds. All existing, recent FPGA accelerators also target outdated, less accurate basecallers that are based on Hidden Markov Models~\cite{rang2018} due to their use of fixed point precision required by most traditional FPGAs and lower computational complexity compared to modern basecallers. We compare CiMBA against a quantized Dorado Fast network basecalling POD5 files, \hll{providing maximum performance, demonstrating favorable performance} in terms of bps/W and bps/mm$^2$ at a fraction of the area/power consumption.

\textit{C. Basecalling-free analysis --} \hll{Rather than performing basecalling, several works~\cite{wu2022, kovaka2021, 2021zhang, cavlak2024, sadasivan2023, gamaarachchi2020, spangenberg2023} analyze raw signals directly. Such strategies benefit a limited number of applications, such as exclusion of non-interesting sequences from the sequencing run~\cite{loose2016}. Unfortunately, such strategies suffer from limited (80-90\%) accuracy~\cite{2021zhang,kovaka2021}, i.e., 10-20\% of useful reads may be excluded.}

By performing basecalling on-chip, CiMBA enables the full range of downstream analyses to be performed without the requisite communication and computation overhead. Further, previous works such as TargetCall~\cite{cavlak2024} have demonstrated that a small, lower accuracy basecaller can be used as an effective "read until" mechanism. In this sense, \hll{a scaled down CiMBA/AL-Dorado system could be implemented to prevent unnecessary sequencing before sending raw data} to the full AL-Dorado network or off-device for high accuracy basecalling.\\

\textit{D. Network optimizations --} \hll{Recent works try to reduce memory footprint and computational complexity via (1) network fixed-point network quantization}~\cite{wu2019, wu2022, hammad2021}, floating point precision with smaller bit-widths~\cite{lou2020}, or mixed precision~\cite{singh2024,lou2018}, and (2) altering skip connections that connect the output of one layer to the input of another nonadjacent layer~\cite{singh2024, weng2023}. \hll{ONT also offers fixed-point INT8 quantization for its Dorado-Fast network.} Other works train basecalling models using species-specific data to improve inference accuracy for that particular species~\cite{ferguson2022}. Such works are orthogonal to CiMBA and can be adapted to further improve CiMBA's performance after carefully examining their benefit for analog inference.

\textit{E. Efficient data representation --} \hll{Nanopore basecalling's FAST5 raw signal storage format is usually 10x larger in size compared to the output of basecalling,} which pressures the compute system with large IO costs and prevents efficient use of parallel CPU resources. More efficient data representation methods are under research in both academia (SLOW5~\cite{gamaarachchi2022}) and industry (POD5 from ONT~\cite{kovaka2023}). Further, no data representation alleviates the communication bottleneck between the sequencer and the workstation before compression can occur. 

By performing on-device basecalling, CiMBA reduces communication and storage costs by compressing on average 10 float32 amplitude values into a single int8 base value, for a \textgreater40x communication/memory reduction.

\section{Conclusion}
\label{sec:conclusion}

This work introduces CiMBA, the first embedded in-memory basecaller capable of running in real-time . CiMBA utilizes a flexible, mesh-based CiM architecture capable of supporting a variety of basecalling networks. Its specialized LookAround decoder enables continuous streaming of inferred bases, permitting deeply pipelined genome analyses. With the co-designed AL-Dorado network, CiMBA reduces communication and storage overhead by 43$\times$ (4.37$\times$) and outperforms the Jetson Xavier AGX embedded GPU by 2$\times$/17$\times$/27$\times$ in terms of throughput, throughput/W, and throughput/mm$^2$, respectively. AL-Dorado is optimized for CiM architectures such as CiMBA, and will be further developed to improve performance and integrate downstream functionality, boosting increased sequencing portability and further opening the door of genomics to new domains.

\bibliographystyle{IEEEtran}
\bibliography{refs}
\vskip -1\baselineskip plus -1fil
\begin{IEEEbiography}[{\includegraphics[width=1in,height=1.25in,clip,keepaspectratio]{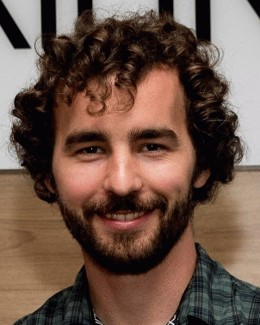}}]{William Andrew Simon} is a research staff member at IBM Research Zurich in the In-Memory Computing group. He received his M.Sc. and Ph.D. degrees in Electrical Engineering at the Swiss Federal Institute of Technology, Lausanne (EPFL). His interests include computer architecture simulation and systems for analog and digital in-memory computing and non-Von Neumann architectures, as well as developing applications for said architectures in various domains including language, vision, and genomics processing.\end{IEEEbiography}
\vskip -1\baselineskip plus -1fil
\begin{IEEEbiography}[{\includegraphics[width=1in,height=1.25in,clip,keepaspectratio]{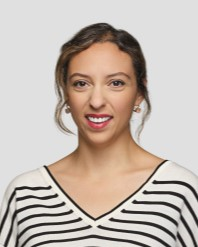}}]{Irem Boybat} is a staff research scientist at IBM Research Zurich in the In-Memory Computing group. She received her Ph.D. degree at EPFL, Lausanne. Her research interests include in-memory computing for AI, application-hardware co-design, and model deployment strategies. She was a co-recipient of the 2018 IBM Pat Goldberg Memorial Best Paper Award and 2020 EPFL PhD Thesis Distinction in Electrical Engineering.\end{IEEEbiography}
\vskip -1\baselineskip plus -1fil
\begin{IEEEbiography}[{\includegraphics[width=1in,height=1.25in,clip,keepaspectratio]{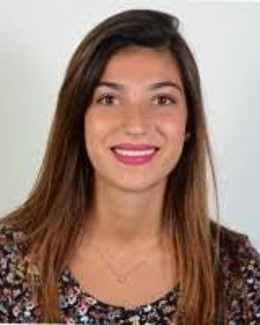}}]{Elena Ferro} is a Predoctoral Researcher in the In-Memory Computing group at IBM Research Zurich. Elena earned her Bachelor's degree in Physical Engineering from Politecnico di Torino, studied Micro- and Nanotechnologies for Integrated Systems and received a double M.Sc. Degree from Politecnico di Torino and Grenoble INP, and a joint degree from the EPFL, Lausanne. Elena's research is primarily centered around digital hardware design and architectures for in-memory computing systems, with a specific focus on applications within the field of AI.\end{IEEEbiography}

\begin{IEEEbiography}[{\includegraphics[width=1in,height=1.25in,clip,keepaspectratio]{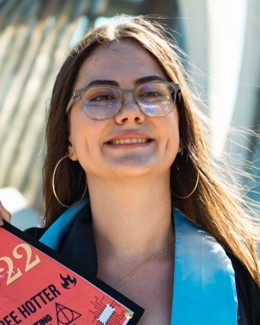}}]{Riselda Kodra} is a M.Sc. student in Electrical Engineering at EPFL, Lausanne. She completed her undergraduate studies in Electronics and Communication at Istanbul Technical University, Turkey. Riselda has worked as a research assistant at the Embedded Systems Laboratory, EPFL where she has gained experience in areas including Near Memory Computing and full system simulators. Her research interests include digital design, memory technologies, VLSI, and AI in biomedical applications.\end{IEEEbiography}
\vskip -1\baselineskip plus -1fil
\begin{IEEEbiography}[{\includegraphics[width=1in,height=1.25in,clip,keepaspectratio]{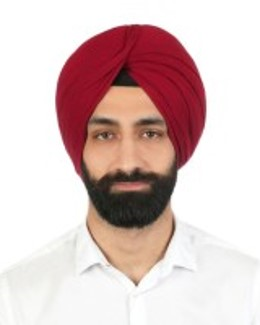}}]{Gagandeep Singh} is a research scientist at Advanced Micro Devices, Inc (AMD). He received a joint M.Sc. degree in integrated circuit design from Technische Universität München, Germany, and Nanyang Technological University, Singapore, and his Ph.D. degree from Technische Universiteit Eindhoven, Netherlands. He was a Predoctoral Researcher with IBM Research Zurich, has worked with Oracle, India, and IMEC, Belgium, and was a Senior Researcher at ETH Zurich. He research interests include computer architecture, FPGA acceleration, processing-in-memory, bioinformatics, and machine learning.\end{IEEEbiography}
\vskip -1\baselineskip plus -1fil
\begin{IEEEbiography}[{\includegraphics[width=1in,height=1.25in,clip,keepaspectratio]{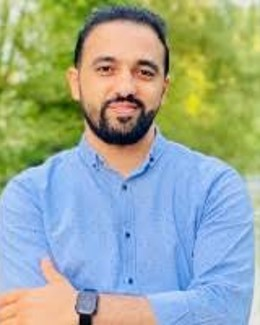}}]{Mohammed Alser} received his Ph.D. in Computer Engineering from Bilkent University. He is an Assistant Professor at Georgia State University. He has previously worked at SAFARI (ETH Zurich), ZarLab (UCLA), CfAED (TU Dresden), and PETRONAS, and has received several international awards, including the ETH Zurich Exceptional Performance Award for two consecutive years, the IEEE Turkey Doctoral Dissertation Award, the Yasser Arafat award, the TÜBITAK doctoral fellowship, and the HiPEAC Collaboration Grant. His primary research interests incorporate several aspects of bioinformatics, metagenomics, computational genomics, and computer architecture.
\end{IEEEbiography}
\vskip -1\baselineskip plus -1fil
\begin{IEEEbiography}[{\includegraphics[width=1in,height=1.25in,clip,keepaspectratio]{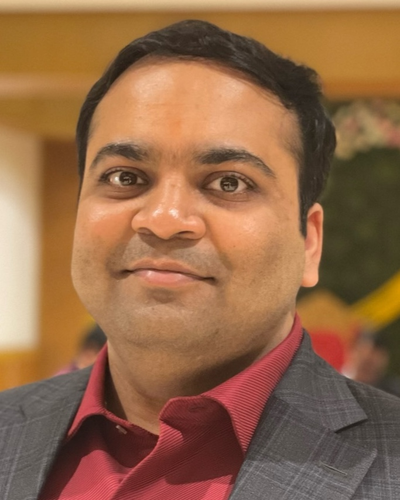}}]{Shubham Jain} is a staff research scientist at IBM Research Yorktown Heights. He received Ph.D. in Electrical and Computer Engineering from Purdue University. Previously, he worked as a design engineer Qualcomm. His primary research interests include AI hardware architecture and Compilers, In-memory computing, and approximate computing. He has published one book chapter and over 27 journal and conference papers, holds 7 US patents, and has 9 pending patent applications. He serves on the Technical Program Committee of the DAC and DATE. He received the Mitacs Globalink scholarship, the Andrews Fellowship from Purdue University, the A. Richard Newton Young Student Fellowship from DAC, and an outstanding TPC member award from DAC. His research has received the DAC Best Technical Paper award, the DATE Best Paper nomination, and a best-in-session award in TECHCON.
\end{IEEEbiography}
\vskip -1\baselineskip plus -1fil
\begin{IEEEbiography}[{\includegraphics[width=1in,height=1.25in,clip,keepaspectratio]{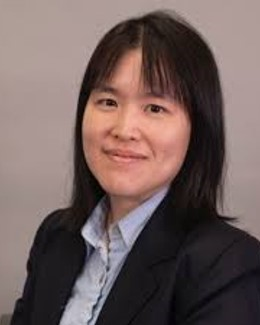}}]{Hsinyu Tsai} received her Ph.D. from the Electrical Engineering and Computer Science department at MIT, USA. She joined the Nanofabrication and Electron Beam Lithography group at the IBM T.J. Watson Research Center as Research Staff Member and developed directed self-assembly (DSA) lithography for finFETs, serving as the manager of the Advanced Lithography group in the Microelectronics Research Laboratory (MRL). She currently works in the Almaden Research Center in San Jose, CA, as a Principal Research Staff Member and manager of the Analog AI group, which has published multiple Nature papers and fabricated two PCM inference chips, highlighted at VLSI in 2021.\end{IEEEbiography}
\vskip -1\baselineskip plus -1fil
\begin{IEEEbiography}[{\includegraphics[width=1in,height=1.25in,clip,keepaspectratio]{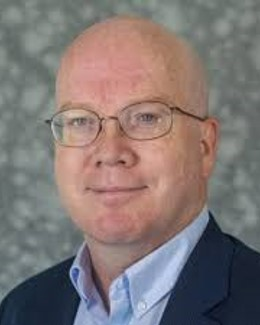}}]{Geoffrey W. Burr} received his Ph.D. in Electrical Engineering from the California Institute of Technology. Since then, Dr. Burr has worked at IBM-Almaden, San Jose, California, where he is a Distinguished Research Scientist, in a number of diverse areas, including holographic data storage, photon echoes, computational electromagnetics, nanophotonics, computational lithography, phase-change memory, storage class memory, and novel access devices based on MIEC materials. Dr. Burr's current research interests involve AI/ML acceleration using non-volatile memory. He is an IEEE Fellow and member of MRS, SPIE, OSA, Tau Beta Pi, Eta Kappa Nu, and the Institute of Physics (IOP).\end{IEEEbiography}
\vskip -1\baselineskip plus -1fil
\begin{IEEEbiography}[{\includegraphics[width=1in,height=1.25in,clip,keepaspectratio]{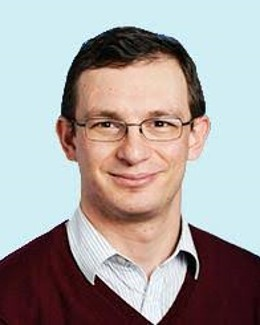}}]{Onur Mutlu} is a Professor of Computer Science at ETH Zurich. He is also a faculty member at Carnegie Mellon University, where he previously held the Strecker Early Career Professorship. His current broader research interests are in computer architecture, systems, hardware security, and bioinformatics. A variety of techniques he, along with his group and collaborators, has invented over the years have influenced industry and have been employed in commercial microprocessors and memory/storage systems. He obtained his Ph.D. and M.Sc. in ECE from the University of Texas at Austin. He started the Computer Architecture Group at Microsoft Research, and held various product and research positions at Intel Corporation, Advanced Micro Devices, VMware, and Google. He received the IEEE Computer Society Edward J. McCluskey Technical Achievement Award, the ACM SIGARCH Maurice Wilkes Award, the inaugural IEEE Computer Society Young Computer Architect Award, the inaugural Intel Early Career Faculty Award, US National Science Foundation CAREER Award, Carnegie Mellon University Ladd Research Award, faculty partnership awards from various companies, and a healthy number of best paper or "Top Pick" paper recognitions at various computer systems, architecture, and hardware security venues. He is an ACM Fellow, IEEE Fellow, and an elected member of the Academy of Europe (Academia Europaea).\end{IEEEbiography}
\vskip -1\baselineskip plus -1fil
\begin{IEEEbiography}[{\includegraphics[width=1in,height=1.25in,clip,keepaspectratio]{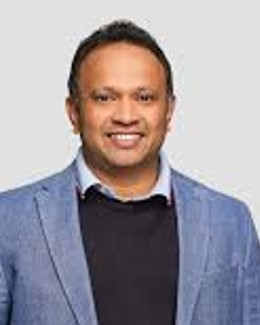}}]{Abu Sebastian} is a distinguished scientist and manager at IBM Research Zurich. He received his M.Sc. and Ph.D. degrees in Electrical Engineering (minor in Mathematics) from Iowa State University. Dr. Sebastian is the author/co-author of over 200 publications in peer-reviewed journals/conference proceedings and holds over 80 US patents. He is a co-recipient of the 2009 IEEE Control Systems Technology Award and the 2009 IEEE Transactions on Control Systems Technology Outstanding Paper Award, the 2013 IFAC Mechatronic Systems Young Researcher Award, a European Research Council (ERC) consolidator grant and a ERC Proof-of-concept grant. He is an IBM Master Inventor since 2016. He was named Principal and Distinguished Research Staff Member in 2018 and 2020, respectively. In 2019 he received the Ovshinsky Lectureship Award for his contributions to 'Phase-change materials for cognitive computing'. In 2023, he received the Prof. L. K. Maheshwari Foundation Distinguished Alumnus Award from BITS Pilani, India. In 2023, he was also conferred the title of Visiting Professor in Materials by University of Oxford. He has served on the technical program committees of several conferences including IEDM, AICAS, NVMTS and and has served as an editor/guest editor for Mechatronics, APL Material, Applied Physics Letters, and IEEE Design and Test. He is a Distinguished Lecturer and Fellow of the IEEE.\end{IEEEbiography}

\end{document}